\documentclass[useAMS,usenatbib]{mn2e}
\usepackage{natbib}
\usepackage{amsmath}
\usepackage{url}
\usepackage{longtable}
\usepackage{aas_macros}
\usepackage{amssymb}
\usepackage{graphicx}
\usepackage{deluxetable}

\newcommand{\about}{$\sim\!\!$~}
\newcommand{\kms}{\,km\,s$^{-1}$}

\def\lsim{\hbox{\rlap{\raise 0.425ex\hbox{$<$}}\lower 0.65ex\hbox{$\sim$}}}
\def\gsim{\hbox{\rlap{\raise 0.425ex\hbox{$>$}}\lower 0.65ex\hbox{$\sim$}}}
\def\arcmin{\hbox{$^\prime$}}
\def\arcsec{\hbox{$^{\prime\prime}$}}

\newcommand\ion[2]{#1$\,${\small{#2}}\relax}

\title[Progenitors of Ca-Rich SNe]{Kinematics and Host-Galaxy Properties Suggest a Nuclear Origin for Calcium-Rich Supernova Progenitors}

\author[Foley]{Ryan~J.~Foley$^{1,2}$\thanks{E-mail:rfoley@illinois.edu}\\
$^{1}$Astronomy Department, University of Illinois at Urbana-Champaign, 1002 W.\ Green Street, Urbana, IL 61801, USA\\
$^{2}$Department of Physics, University of Illinois at Urbana-Champaign, 1110 W.\ Green Street, Urbana, IL 61801, USA}

\begin{document}

\date{Accepted  . Received   ; in original form  }
\pagerange{\pageref{firstpage}--\pageref{lastpage}} \pubyear{2013}
\maketitle
\label{firstpage}

\begin{abstract}
  Calcium-rich supernovae (Ca-rich SNe) are peculiar low-luminosity
  SNe~Ib with relatively strong Ca spectral lines at \about 2 months
  after peak brightness.  This class also has an extended projected
  offset distribution, with several members of the class offset from
  their host galaxies by 30 -- 150~kpc.  There is no indication of any
  stellar population at the SN positions.  Using a sample of 13
  Ca-rich SNe, we present kinematic evidence that the progenitors of
  Ca-rich SNe originate near the centers of their host galaxies and
  are kicked to the locations of the SN explosions.  Specifically, SNe
  with small projected offsets have large line-of-sight velocity
  shifts as determined by nebular lines, while those with large
  projected offsets have no significant velocity shifts.  Therefore,
  the velocity shifts must not be primarily the result of the SN
  explosion.  There is an excess of SNe with blueshifted velocity
  shifts within two isophotal radii (5/6 SNe), indicating that the SNe
  are moving away from their host galaxies and redshifted SNe on the
  far sides of their galaxies are selectively missed in SN surveys.
  Additionally, nearly every Ca-rich SN is hosted by a galaxy with
  indications of a recent merger and/or is in a dense environment.  At
  least 6--7 host galaxies also host an AGN, a relatively high
  fraction, again linking the nuclear region of these galaxies to
  explosions occurring tens of kpc away.  We propose a progenitor
  model which fits all current data: The progenitor system for a
  Ca-rich SN is a double white dwarf (WD) system where at least one WD
  has a significant He abundance.  This system, through an interaction
  with a super-massive black hole (SMBH) is ejected from its host
  galaxy and the binary is hardened, significantly reducing the merger
  time.  After 10 -- 100~Myr (on average), the system explodes with a
  large physical offset.  The rate for such events is significantly
  enhanced for galaxies which have undergone recent mergers,
  potentially making Ca-rich SNe new probes of both the galaxy merger
  rate and (binary) SMBH population.
\end{abstract}



\defcitealias{Perets10:05e}{P10}
\defcitealias{Kasliwal12}{K12}

\section{Introduction}\label{s:intro}

For the past century, systematic supernova (SN) searches have
discovered thousands of SNe.  Nearly all of these SNe fall into 3
classes: Type Ia, Type II, and Type Ib/c.  However, over the last
decade with the implementation of large SN searches, we have begun to
discover many astrophysical transients that do not fall into the
well-delineated classes mentioned above.  It turns out that there were
about a dozen new classes of ``exotic'' or ``peculiar'' transients
lurking in the shadows.  These classes include luminous SNe~IIn
\citep[e.g.,][]{Smith07:06gy}, Type I super-luminous SNe
\citep{Quimby11}, kilonovae \citep{Berger13:kilo, Tanvir13}, SNe~Iax
\citep{Foley13:iax}, and SN~2006bt-like SNe \citep{Foley10:06bt}.

One class, the SN~2005E-like SNe \citep[hereafter,
\citetalias{Perets10:05e}]{Perets10:05e}, also known as ``Calcium-rich
SNe'' \citep[or simply ``Ca-rich SNe'';][hereafter,
\citetalias{Kasliwal12}]{Filippenko03:carich, Kasliwal12}, are
particularly interesting.  These SNe are technically of Type Ib,
having distinct He lines in their spectra near maximum brightness.
However, they are less luminous and faster fading than normal SNe~Ib
\citepalias{Perets10:05e, Kasliwal12} and are often found in
early-type galaxies (\citetalias{Perets10:05e}; \citealt{Lyman13}).
Additionally, several members of the class, including SN~2005E have
large projected offsets ($>$10~kpc and up to at least 150~kpc) from
their host galaxies (\citetalias{Perets10:05e, Kasliwal12};
\citealt{Yuan13, Lyman14}; this work).

Relative to other SNe~Ib, Ca-rich SNe also have a distinct
spectroscopic evolution.  After about 2 months, the SNe show strong
forbidden lines, indicating that the ejecta are at that point mostly
optically thin.  This is a much faster transition than for most
SNe~Ib.  This ``nebular'' spectrum is also distinct, showing extremely
strong [\ion{Ca}{II}] $\lambda\lambda7291$, 7324 emission relative to
that of [\ion{O}{I}] $\lambda\lambda6300$, 6363, giving the name to
the class \citep{Filippenko03:carich}.  Modeling of the ejecta also
indicates that the mass-fraction of Ca in the ejecta is also extremely
high \citepalias[\about 1/3;][]{Perets10:05e}.

\citetalias{Perets10:05e} examined the rate of Ca-rich SNe in the Lick
Observatory Supernova Search \citep[LOSS;][]{Filippenko01}, finding
that they occur at $7 \pm 5$\% the rate of SNe~Ia.
\citetalias{Kasliwal12} found a rate $>$2.3\% that of SNe~Ia for the
Palomar Transient Factory (PTF) sample.  Therefore, Ca-rich SNe are
somewhat common events and cannot come from extremely rare progenitor
scenarios.

Since most Ca-rich SNe have early-type hosts
(\citetalias{Perets10:05e}; \citealt{Lyman13}), a massive star origin
is unlikely.  Moreover, there is no indication of star formation at
the position of the SNe in deep pre- or post-explosion imaging
(\citetalias{Perets10:05e}; \citealt{Perets11};
\citetalias{Kasliwal12}; \citealt{Lyman14}).

Because of the large offset distribution, it has been suggested that
Ca-rich SNe occur in dwarf galaxies (\citetalias{Kasliwal12};
\citealt{Yuan13}) or globular clusters \citep{Yuan13}; however, the
deep limits on any stellar light at the position of the SNe rule out
such possibilities \citep{Lyman14}.  These observations require that
the progenitors of Ca-rich SNe be born elsewhere and travel a
significant distance to where they explode \citep{Lyman14}.  With this
determination, \citet{Lyman14} suggested that Ca-rich SN progenitors
were neutron star (NS) -- white dwarf (WD) binary systems which are
kicked by the SN that created the NS and then undergo a merger after
traveling far from their birth site.

In this manuscript, we present a kinematic and host-galaxy study of
Ca-rich SNe.  By examining the velocity shift distribution, the
projected offset distribution, and the angle offset distribution, as
well as the host-galaxy properties of the sample, we constrain the
progenitor systems and origin of Ca-rich SNe.  In
Sections~\ref{s:sample} and \ref{s:gal}, we define our SN sample and
discuss their host-galaxy properties.  In Sections~\ref{s:off} and
\ref{s:shifts}, we measure the projected offset and velocity shift
distributions for our sample.  In Section~\ref{s:anal}, we analyze
these data.  We present a basic progenitor model in
Section~\ref{s:model}.  We discuss our results and summarize our
conclusions in Section~\ref{s:disc}.


\section{Supernova Sample}\label{s:sample}

\setcounter{figure}{0}
\begin{figure*}
\begin{center}
\includegraphics[angle=0,width=6.4in]{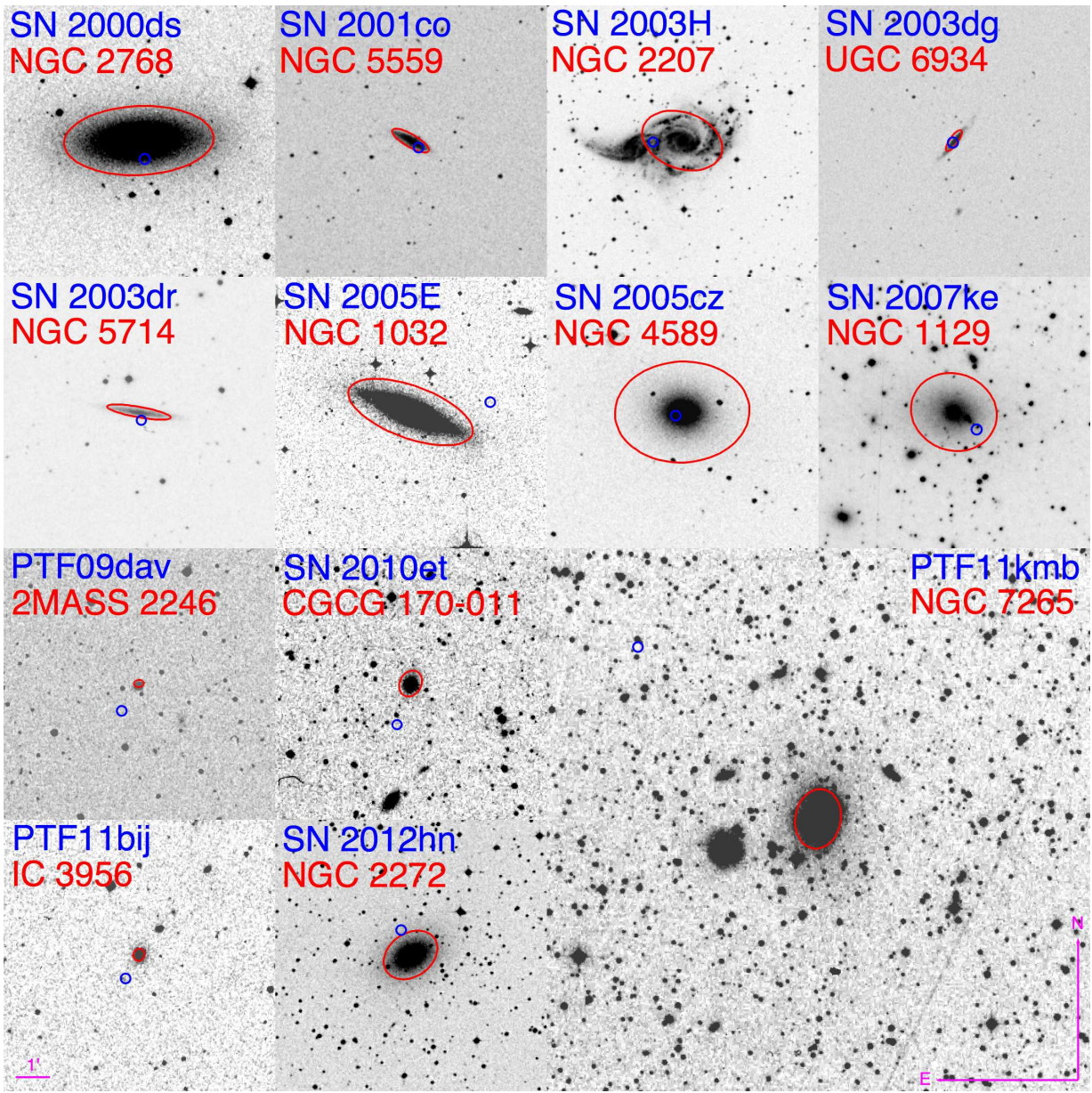}
\caption{DSS images of Ca-rich SN host galaxies, with each centered in
  individual panels.  Each panel is $8^{\prime} \times 8^{\prime}$
  except for PTF11kmb, which is $16^{\prime} \times 16^{\prime}$.  All
  panels have the same scale and are aligned with North up and East to
  the left.  Each SN position is marked with a blue circle.  An
  ellipse at twice the isophotal radius is shown in
  red.}\label{f:montage}
\end{center}
\end{figure*}

Our sample consists of all known (and published) Ca-rich SNe.  We
begin with the \citet{Lyman14} sample of Ca-rich SNe.  This sample
includes 12 SNe originally reported in other works
(\citetalias{Perets10:05e}; \citealt{Kawabata10, Sullivan11:09dav};
\citetalias{Kasliwal12}; \citealt{Valenti14}).  We include PTF09dav in
this analysis, although at maximum light, it had much lower ejecta
velocities than other members \citep{Sullivan11:09dav} and uniquely
displays hydrogen emission at late-times \citepalias{Kasliwal12}.

To this sample, we add PTF11kmb.  PTF11kmb was discovered on 24.24
August 2011 (all times are UT) by PTF \citep{Gal-Yam11}.  A Keck/LRIS
spectrum obtained 28 August 2011 was used to classify PTF11kmb as a SN
Ib \citep{Gal-Yam11}.  We use this spectrum as well as a Keck/LRIS
spectrum from 26.37 November 2011 to classify PTF11kmb as a Ca-rich SN
(Appendix~\ref{a:11kmb}).

Basic host-galaxy information was obtained through the NASA/IPAC
Extragalactic Database (NED) and is presented in Table~\ref{t:host}.
Four SNe have somewhat ambiguous hosts, and we discuss each below.

SN~2003H occurred between two merging galaxies, NGC~2207 and IC~2163.
Their recession velocities differ by $24 \pm 25$~km~s$^{-1}$, and this
difference does not affect our results.  SN~2003H is offset by 8.73
and 5.75~kpc from NGC~2207 and IC~2163, respectively, and the
difference in offset does not affect any results.

SN~2007ke occurred in a cluster environment (AWM~7) near the brightest
member, NGC~1129.  It is also near a group member, MCG+07-07-003.  The
difference in their recession velocities is $227 \pm 23$~km~s$^{-1}$,
with NGC~1129 having the larger redshift.  This difference does not
affect any results, but is noteworthy.  The SN was offset by 16.71 and
8.20~kpc from NGC~1129 and MCG+07-07-003, respectively.  Again, this
difference does not affect any results.  Given the relative sizes of
these galaxies and large offset from MCG+07-07-003, NGC~1129 is a
reasonable choice for the host galaxy.  Nonetheless, given the cluster
environment, it is possible that SN~2007ke originated from a faint
cluster member or was formed in the intracluster medium.

SN~2010et occurred 37.6 -- 69.2~kpc from three galaxies at redshifts
consistent with the SN redshift.  The closest galaxy, CGCG~170-011 is
also relatively large.  Using the NED reported major and minor 2MASS
$K_{s}$ isophotal axes (with a reference value of $K_{s} =
20.0$~mag~arcsec$^{-2}$) and position angle, SN~2010et was offset 5.5
isophotal radii from CGCG~170-011.  SN~2010et is 69.2~kpc and 16.5
isophotal radii from the other nearby, large galaxy, CGCG~170-010.  A
third galaxy, SDSS~J171650.20+313234.4, is 46.9~kpc and is too small
and faint for a 2MASS isophotal measurement.  Based on offset, the
most likely host is CGCG~170-011, and we assume this for the rest of
the analysis.  \citetalias{Kasliwal12} determined that no source
exists at the position of SN~2010et with $M_{R} < -12.1$~mag, but this
does not completely rule out a possible dwarf galaxy as the host of
SN~2012et.

Finally, PTF11kmb is far from any galaxy.  PTF11kmb has a redshift of
roughly $z = 0.017$ as determined by SN features \citep[and confirmed
in our analysis]{Gal-Yam11}.  The closest galaxy listed in NED is
4.2~arcmin away and has no redshift information.  However, if it is at
the distance of PTF11kmb, it would be offset by \about 80~kpc.  The
three closest galaxies with redshifts are 2MASX J22224094+3613514
(offset by 4.5~arcmin, 83.0~kpc, and 36.6 isophotal radii; $cz = 4500
\pm 25$~km~s$^{-1}$), UGC~12007 (offset by 6.5~arcmin, 127.3~kpc, and
15.0 isophotal radii; $cz = 4829 \pm 33$~km~s$^{-1}$), and NGC~7265
(offset by 7.3~arcmin, 150.0~kpc, and 10.2 isophotal radii; $cz = 5083
\pm 26$~km~s$^{-1}$).

While NGC~7265 is very far physically from PTF11kmb, it is the closest
in terms of isophotal radii and has the most similar redshift to that
derived from the SN.  We therefore adopt NGC~7265 as the host of
PTF11kmb for this analysis.

All spectra (except for that of PTF11kmb) were previously published
(\citealt{Kawabata10}; \citetalias{Perets10:05e};
\citealt{Sullivan11:09dav}; \citetalias{Kasliwal12};
\citealt{Valenti14}).  We obtained some of these data through WISERep
\citep{Yaron12}.

Almost all SNe in our sample have only one late-time spectrum.  For
the few SNe with multiple late-time spectra, we primarily analyze the
latest high-quality spectrum available.  The spectra typically had
phases of \about 2 months after maximum brightness
\citepalias[e.g.,][]{Kasliwal12}.  At these phases, and for all
spectra examined here, Ca-rich SNe have strong [\ion{Ca}{II}]
$\lambda\lambda7291$, 7324 emission, especially relative to that of
[\ion{O}{I}] $\lambda\lambda6300$, 6363.


\section{Host-galaxy Properties}\label{s:gal}

\subsection{Evidence for Recent Host-galaxy Mergers}\label{ss:merger}

In Figure~\ref{f:montage}, we present Digitized Sky Survey (DSS)
images of the host galaxies of the Ca-rich SN sample.  The host
galaxies of Ca-rich SNe are clearly atypical as previously noted
(\citetalias{Perets10:05e}; \citealt{Lyman13}); the sample is highly
skewed to early-type galaxies.

However, previous studies had not noted the strong evidence for recent
host-galaxy mergers.  In fact, nearly every Ca-rich SN host galaxy
shows some indication of a recent merger and/or is in a very dense
environment where the likelihood of recent galactic mergers is much
larger than in the field.

Of the 13 host galaxies, one is clearly interacting (NGC~2207) and one
is a disturbed spiral
\citep[2MASS~J22465295+2138221;][]{Sullivan11:09dav}.  NGC~1129 has a
disturbed morphology indicative of a recent merger \citep{Peletier90}.
There are also four S0 galaxies (IC~3956, NGC~1032, NGC~2272, and
NGC~7265), which are often thought to be the result of recent mergers
\citep[e.g.,][]{Moore99} or which may have been ``harassed''
\citep[e.g.,][]{Moore98}.  From these tracers alone, 7/13 Ca-rich host
galaxies have some indication of a recent merger.

Additionally, at least 8/13 host galaxies are in a group or cluster
environment.  This is an exceedingly high fraction, even for
elliptical galaxies.  Moreover, 7/13 host galaxies are either the
brightest group galaxy (BGG) or brightest cluster galaxy (BCG).  As
these galaxies tend to sit at or near the center of the group/cluster
potential, they are more likely to have had recent mergers than
typical cluster members.

The relative rate of Ca-rich SNe to SNe~Ia is roughly 7\%
\citepalias{Perets10:05e}.  Using the relative fraction of ``recent
merger'' host galaxies for the two groups, the Ca-rich SNe would have
a relative rate of only $2 \pm 2$\% that of SNe~Ia in galaxies with no
indications of recent mergers and $11 \pm 8$\% that of the SN~Ia rate
in galaxies with some indication of a recent merger.

In total, 11/13 Ca-rich SN host galaxies either have some evidence for
a ongoing/recent merger or are in dense environments.  Using the LOSS
SN~Ia sample \citep{Leaman11}, we find that 52\% of SN~Ia host
galaxies have similar host properties.  Using binomial statistics,
there is only a 1.6\% chance that SNe~Ia and Ca-rich SNe come from
similar host populations in terms of their ``recent merger''
properties.  Since Ca-rich SN and SN~Ia host galaxies have
indistinguishable morphology and star-formation rate distributions
(\citetalias{Perets10:05e}; \citealt{Lyman13}), the merger/environment
is likely a significant factor in the relative rates.  The galaxies in
rich environments or that have had recent mergers have an enhanced
rate of Ca-rich SNe relative to that of SNe~Ia.

\subsection{A High Occurrence of AGN}\label{ss:agn}

\begin{figure}
\begin{center}
\includegraphics[angle=90,width=3.2in]{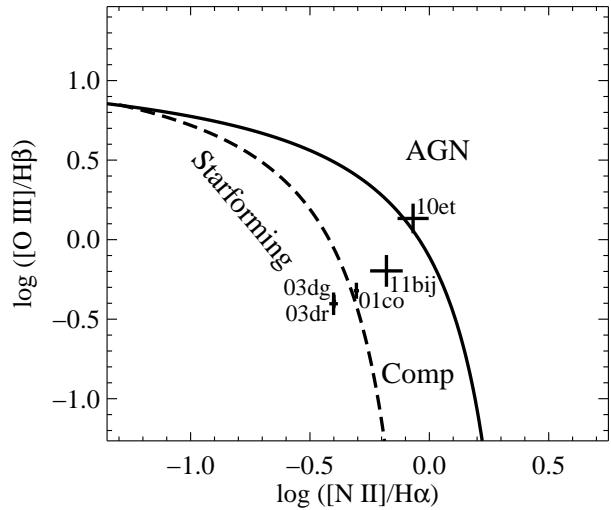}
\caption{Baldwin-Phillips-Terlevich (BPT) diagram \citep{Baldwin81}
  for the Ca-rich SN host galaxies with SDSS spectra.  The curves
  delineating the star-forming region (below the dashed line), the AGN
  region (above the solid line) and the ``composite'' region (between
  the dashed and solid lines) are taken from \citet{Kewley01} and
  \citet{Kauffmann03} as implemented in
  \citet{Kewley06}.}\label{f:bpt}
\end{center}
\end{figure}

In addition to the different indications of merger activity, several
host galaxies also host an active galactic nucleus (AGN).  Previously
identified AGN include NGC~2207 \citep{Kaufman12}, NGC~2768
\citep{Veron-Cetty06}, NGC~4589 \citep{Ho97:params, Nagar05}, and
IC~3956 \citep[e.g.,][]{Liu11}.  Additionally, NGC~1032 has a
radio-to-far infrared flux ratio consistent with an AGN
\citep{Drake03}.  We also examined the Sloan Digital Sky Survey (SDSS)
emission line quantities for the host galaxies observed by SDSS
\citep[e.g.,][]{Tremonti04}.  For these five galaxies, we identify
CGCG~170-011 as a likely AGN, confirm that IC~3956 has a ``composite''
spectrum indicative of AGN activity, and identify NGC~5559 as having a
likely composite spectrum (Figure~\ref{f:bpt}).

At least 6--7 (depending on the classification of NGC~1032) of the 13
Ca-rich host galaxies also host an AGN.  This is a large fraction
considering that this is an incomplete survey of Ca-rich host
galaxies.  Although the total AGN fraction may be as high as \about
43\% when the lowest luminosity galaxies and AGN are included
\citep[e.g.,][]{Ho08}, this fraction can be lower for specific galaxy
samples.  For instance, an SDSS spectroscopic sample of $>$120,000
galaxies found that 18\% of all galaxies hosted an AGN as determined
from line diagnostics \citep{Kauffmann03}.  For the host galaxies with
SDSS spectra, 3/5 galaxies also host an AGN.  Using binomial
statistics for this subsample with the SDSS occurrence rate, there is
a 4.7\% chance that the Ca-rich host galaxies are typical.  If we
expand this to the larger, incomplete sample, we find a $<$0.4--2.0\%
chance (depending on if 6 or 7 galaxies host an AGN).  However, we
note that if we use a 43\% occurrence rate, these differences are not
significant for any subsample.  For such a rate, $\ge$9/13 galaxies
must host an AGN for a significant difference.

There is likely a connection between the nuclear regions of these
galaxies and the SNe offset up to 150~kpc in projection.  Since the
typical duty cycle of an AGN is \about $10^{7}$~years, the information
from the nucleus to SN location would have to travel at a very high
speed of $\gtrsim$1000~\kms.


\section{Galactic Offsets}\label{s:off}

Ca-rich SNe have been found very far from their host galaxies
\citepalias[e.g.,][]{Perets10:05e, Kasliwal12}.  However, this is not
exclusively the case.  For instance, SN~2003dg was offset 4.1\arcsec\
from its host galaxy corresponding to a projected distance of only
1.7~kpc.

Using NED values for the angular size scale and offsets between the
SNe and host-galaxy, we determine a projected offset for each SN.  The
typical uncertainty in the offsets is \about0.1\arcsec, corresponding
to projected offset uncertainties of only 0.01 -- 0.07~kpc.  All SN
properties, including those relative to their host galaxies, are
presented in Table~\ref{t:sn}.

In Figure~\ref{f:montage}, we present Digitized Sky Survey (DSS) images
of the host galaxies of the Ca-rich SN sample.  In each image, we
display an ellipse corresponding to two isophotal radii for each
galaxy and the SN position.

To determine the isophotal radial offset for each SN, we use the 2MASS
$K_{s}$ isophotal elliptical parameters (with a reference value of
$K_{s} = 20.0$~mag~arcsec$^{-2}$) for each galaxy.  For each SN
position, we determine a multiplicative factor such that an ellipse
with the same axis ratio, position angle, and center will intersect
with the SN position.  This multiplicative factor is the number of
isophotal radii at which the SN is offset.

There are many sources for isophotal parameters, but 2MASS provides a
large and homogeneous source.  However, these parameters may have
somewhat large errors for particular galaxies.  The most suspect value
is the position angle for NGC~2207.  The 2MASS position angle is
$70^{\circ}$, while other sources have values of $97^{\circ}$ and
$141^{\circ}$.  It appears that these differences are caused by the
different position angles of the bulge/bar and the spiral arms.
Nonetheless, NGC~2207 has an axis ratio of 0.68, so the exact angle
has only a marginal affect on the measured isophotal radial offset for
SN~2003H.

Some galaxies are highly inclined, which is important for
understanding the SN offsets.  For instance, SN~2003dr is offset only
2.7~kpc (in projection) from its host galaxy nucleus, but it is offset
nearly perpendicular to its nearly edge-on host galaxy (minor axis
relative to major axis of 0.16).  Thus SN~2003dr is offset by 2.9
isophotal radii from the nucleus of its host galaxy.

\subsection{Galaxy Targeted  SN Search Efficiency}\label{ss:target}

Some Ca-rich SNe have extremely large angular offsets.  For instance,
SN~2005E and PTF11kmb are offset 2.3$^{\prime}$ and 7.3$^{\prime}$
from their host galaxies, respectively.  Galaxy-targeted SN searches,
such as LOSS, could potentially miss several Ca-rich SNe because they
are offset beyond the field-of-view (FOV) of the camera.  This is
particularly worrisome since PTF, which runs an untargeted search and
has a large FOV camera, has only discovered (and announced) Ca-rich
SNe that have a projected distance of at least 34~kpc and up to
150~kpc.

To better understand the efficiency of the LOSS search at finding
Ca-rich SNe, we perform a simple calculation.  The KAIT camera has a
$7.8^{\prime} \times 7.8^{\prime}$ FOV.  By assuming that a targeted
galaxy is centered in the middle of the FOV, we can determine the
fraction of projected offsets observed for a galaxy at a given
distance.

\begin{figure}
\begin{center}
\includegraphics[angle=90,width=3.2in]{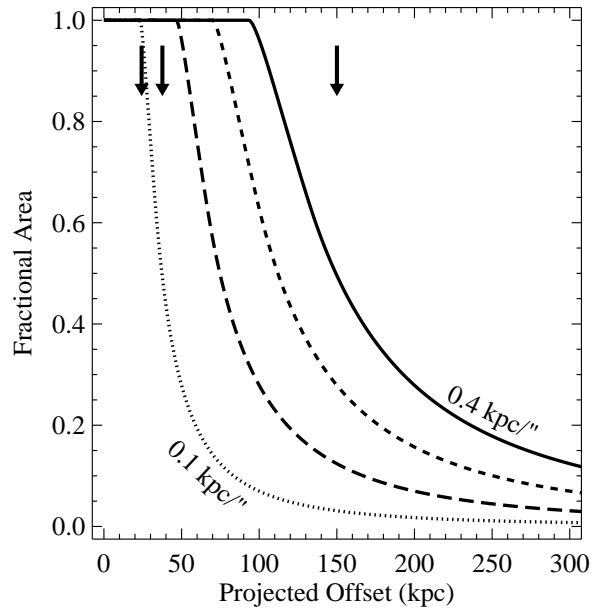}
\caption{Fraction of the region within a certain projected offset from
  a particular galaxy observed by KAIT.  The dotted, long-dashed,
  short-dashed, and solid lines correspond to scales of 0.1, 0.2, 0.3,
  and 0.4~kpc/\arcsec, respectively.  The arrows represent, from
  smaller to larger offsets, the projected offsets of SN~2005E,
  PTF~11bij, and PTF11kmb, respectively.}\label{f:kait}
\end{center}
\end{figure}

Figure~\ref{f:kait} displays the results from this calculation.  For
very nearby galaxies, where the scale is 0.1~kpc/\arcsec, KAIT
observes 10, 50, and 100\% of the area within 83, 37, and 23~kpc,
respectively.  Since many Ca-rich SNe have peak absolute magnitudes of
$M \approx -16$ \citepalias{Perets10:05e, Kasliwal12}, KAIT will be
magnitude limited for Ca-rich SNe beyond \about 80~Mpc.  At this
distance, the scale is \about 0.4~kpc/\arcsec.  For this scale, KAIT
observes 10, 50, and 100\% of the area within 333, 149, and 93~kpc,
respectively.  A typical Ca-rich host galaxy within the KAIT sample
may have \about 0.3~kpc/\arcsec, where KAIT observes 10, 50, and 100\%
of the area within 250, 112, and 70~kpc, respectively.

The farthest offset in the Ca-rich SN sample is PTF11kmb with an
offset of 150~kpc.  Of the remaining PTF-discovered Ca-rich SNe, the
median offset is 37.6~kpc, and therefore KAIT should have been able to
discover such SNe in nearly all of their targeted galaxies.  SN~2005E,
which was discovered by LOSS \citep{Graham05}, was offset by 24.3~kpc.
In fact, all Ca-rich SNe, with the exception of PTF11kmb, are offset
less than 3.9${\arcmin}$, meaning that they would land within the KAIT
FOV.

Although KAIT will not be able to discover SNe with extremely large
offsets, especially for the closest galaxies, it should be able to
sample the bulk of the population discovered by PTF.  Although there
are likely systematic effects related to humans identifying a
transient far offset from a host galaxy as a true SN, there is no
technical reason for a lack of Ca-rich SNe with large offsets in the
LOSS sample.

Since LOSS is not clearly missing Ca-rich SNe with large offsets, it
is reasonable to assume that PTF is missing some Ca-rich SNe with
small offsets (or not reporting these objects as Ca-rich SNe).  This
may be caused by the difficulty of detecting faint sources on top of
small galaxies, the difficulty of obtaining high-quality spectra of
faint sources on top of relatively bright galaxies, or a preference to
obtain multi-epoch, high-quality spectra of SNe~Ib with large offsets.

PTF not detecting or not classifying some Ca-rich SNe projected on top
of galaxies may also explain the ``low'' rate for Ca-rich SNe relative
to the KAIT rate measurement \citepalias{Perets10:05e, Kasliwal12}.


\section{Velocity Shifts}\label{s:shifts}

Forbidden lines, which are produced by optically thin material, are
extremely useful for understanding the velocity distribution of SN
ejecta along our line of sight.  In particular, forbidden emission
lines provide the opportunity to measure the bulk velocity offset of
the SN ejecta along our line of sight.

For Ca-rich SNe, there are two optical forbidden line complexes which
may be studied, [\ion{O}{I}] $\lambda\lambda6300$, 6363 and
[\ion{Ca}{II}] $\lambda\lambda7291$, 7324.  As a defining property of
this class, the SNe have much stronger [\ion{Ca}{II}] lines than
[\ion{O}{I}].  As a result, the majority of our sample has weak or
undetectable [\ion{O}{I}] lines.  However, the [\ion{Ca}{II}] lines are
sufficiently strong for proper kinematic measurements in all spectra
used for this analysis.

\subsection{Velocity Measurement Method}

For a single emission line, there are three typical measurements used
to determine the bulk velocity offset: the peak of emission, the
center of a Gaussian fit to the emission profile, and the
emission-weighted velocity.  If a line profile is Gaussian, the same
velocity would be measured from all three methods.  However, if there
is significant skewness to the profile or multiple peaks, the methods
could differ.

For [\ion{Ca}{II}] $\lambda\lambda7291$, 7324 in SN spectra, the
doublet is not typically resolved, and this is the case for Ca-rich
SNe as well.  However, the lines are separated enough such that a
single Gaussian is generally not a good description of the data.
Moreover, the line profiles are generally moderately skewed, making
even the sum of two Gaussian profiles (offset to match the wavelength
offset of the doublet) a poor description of the data.

\begin{figure}
\begin{center}
\includegraphics[angle=90,width=3.2in]{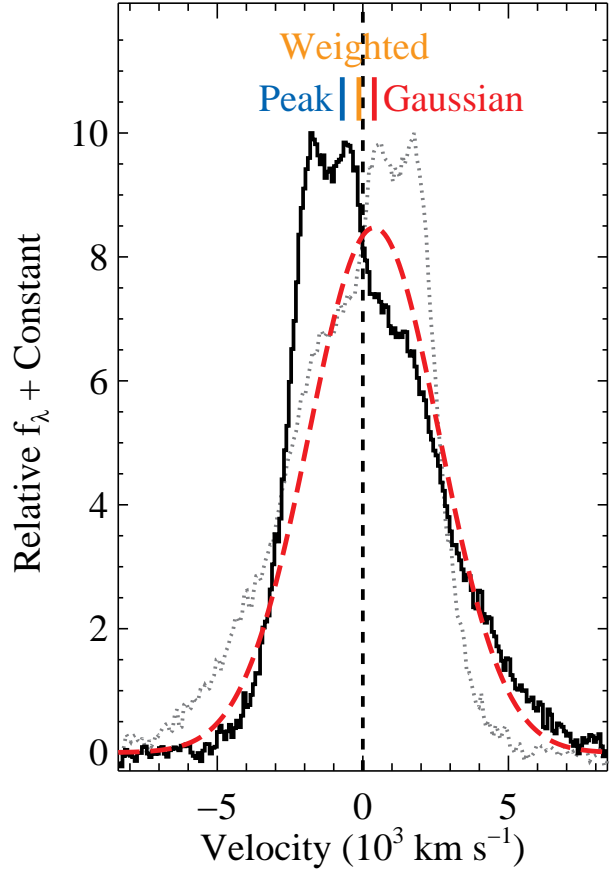}
\caption{Continuum-subtracted [\ion{Ca}{II}] $\lambda\lambda7291$,
  7324 profile for SN~2005E (black curve).  The spectrum is also
  reflected across zero velocity (dotted grey curve) to demonstrate
  the skewness in the tails of the feature.  A best-fit double
  Gaussian profile with rest wavelengths corresponding to those of the
  [\ion{Ca}{II}] doublet is shown as the red dashed curve.  The
  velocity shifts corresponding to the peak of the smoothed line
  profile, the emission-weighted center of the profile, and the shift
  of the double-Gaussian profile are marked with small vertical blue,
  gold, and red lines.}\label{f:profile}
\end{center}
\end{figure}

As an example, Figure~\ref{f:profile} shows the [\ion{Ca}{II}]
$\lambda\lambda7291$, 7324 doublet for SN~2005E.  For all line
profiles, we subtract the underlying continuum by linearly
interpolating across the line profile.  The shape of the profile is
complicated and far from Gaussian.  The profile peaks blueward of the
nominal central wavelength with a large drop in emission around zero
velocity.  Although this has been an indication of dust formation in
SN ejecta \citep[e.g.,][]{Smith08:06jc}, the long tail of redward
emission (extending perhaps 3000~km~s$^{-1}$ further to the red than
the blue) suggests that the line profile cannot be explained simply by
dust reddening.  Instead, the profile is likely indicative of a
complex velocity structure for the SN ejecta.

Clearly a single Gaussian is not an appropriate description of the
[\ion{Ca}{II}] profile for SN~2005E.  We have also attempted to fit the
profile with two Gaussians where their rest-frame wavelength offsets
correspond to the difference in wavelengths for the doublets, and
their widths and heights are constrained to be the same.  These
constraints should result in a perfect match to the data if the
emitting material were described by a simple velocity distribution.
However, SN~2005E has a more complex velocity distribution.  Because
of the poor Gaussian fits for many spectra in our sample, we do not
use this method for our analysis.

A second possibility is to use the peak of emission.  Since most
spectra contain noise at the level that would affect results (or
possibly small scale structure), we smooth the spectra before
determining the peak.  For SN~2005E, the velocity of the peak of
emission is offset from the best-fit Gaussian velocity by
1100~km~s$^{-1}$.  The peak of the emission is a particularly poor
tracer of the entire SN ejecta since it lacks any information about
the shape of the velocity distribution, can be influenced by small
density perturbations in the ejecta, and is strongly affected by noise
in the spectra.  We therefore do not use this method of our analysis.

The final possibility is the emission-weighted velocity.  Relative to
this velocity, half of the emission is blueshifted (and half
redshifted).  This method uses all information from the line profile,
is not affected by the two emission lines, is not significantly
affected by noise, and is not affected by skewness.  Moreover, this
velocity has a specific physical meaning.  If the ejecta are
completely optically thin and each Ca atom in the ejecta has an equal
probability of emitting a photon at these wavelengths, this velocity
will correspond to the systemic line-of-sight velocity.  As this is
clearly the best velocity measurement for determining any
line-of-sight velocity shifts, we use this method for our analysis.

\subsection{Systematic Uncertainties}

While the method chosen to measure of the velocity shift is relatively
robust, we investigate possible systematic uncertainties, which
dominate the uncertainty of the velocity measurement.

To determine the uncertainty, we performed a Monte Carlo simulation
for each spectrum.  For each spectrum, we smoothed the [\ion{Ca}{II}]
line profile and measured the emission-weighted velocity.  Then using
noise properties matched to the data, we simulated 1000 spectra for
each spectrum.  We measured the emission-weighted velocity of each
simulated spectrum, randomly changing the regions used to determine
the continuum.  There was no bias in these measurements, with most
shifts from the noise-free spectrum being one or two pixels.  This
demonstrates the robustness of this method to noisy spectra and with
different choices for the continuum.  We use the standard deviation of
the shifts from the noise-free measurement as the uncertainty in the
emission-weighted velocity.

\begin{figure*}
\begin{center}
\includegraphics[angle=90,width=6.4in]{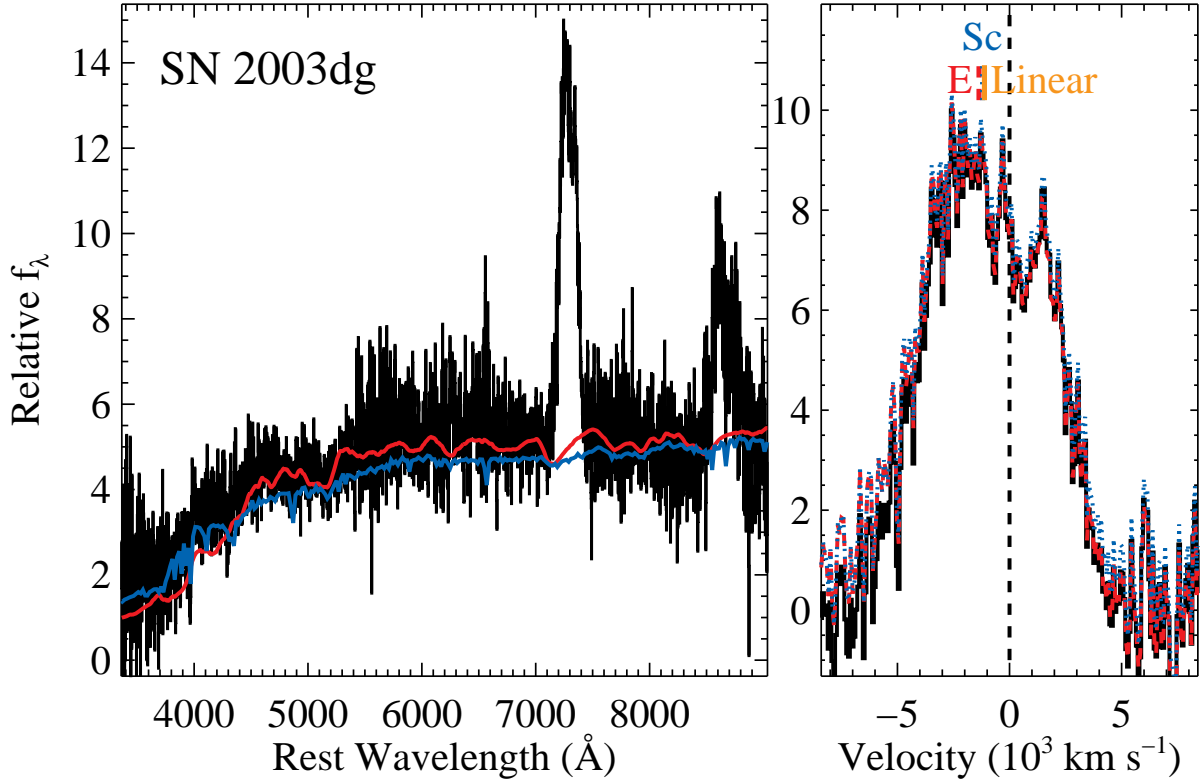}
\caption{{\it Left}: Optical spectrum of SN~2003dg (black).
  Overplotted are template elliptical (red) and reddened ($E(B-V) =
  0.35$~mag) Sc (blue) spectra.  The galaxy spectra near
  [\ion{Ca}{II}] $\lambda\lambda7291$, 7324 is not necessarily linear.
  {\it Right}: Corresponding continuum-subtracted [\ion{Ca}{II}]
  $\lambda\lambda7291$, 7324 profile for SN~2003dg.  The black, red,
  and blue spectra correspond to a linear continuum, elliptical
  galaxy, or reddened Sc galaxy subtracted from the SN spectrum.  The
  resulting velocity shifts are plotted above the profile.  For this
  example, the choice of continuum results in a systematic difference
  in the velocity shift of $-240$ and $-80$~\kms for the elliptical
  and Sc templates, respectively.}\label{f:galcont}
\end{center}
\end{figure*}

There is an additional potential systematic uncertainty related to the
continuum subtraction.  Our method assumes a linear underlying
continuum.  However, if the continuum is dominated by a galaxy
spectrum, a linear continuum is not necessarily appropriate.  An
example of likely host-galaxy contamination for a SN spectrum is shown
in Figure~\ref{f:galcont}, where we show the spectrum of SN~2003dg.

We examine the effect of subtracting a galaxy continuum rather than a
linear continuum.  For this task, we choose an elliptical template
spectrum, which appears to be an excellent match to the continuum for
SN~2003dg, and a reddened Sc template spectrum since its host galaxy
is classified as Scd.  Using the elliptical and Sc template spectra
result in velocity shifts that are systematically 240 and 80~\kms\
bluer than the linear continuum resulting in a velocity shift of
$-1360$ and $-1200$~\kms, respectively, compared to our nominal value
of $-1120$~\kms.

We have examined all spectra, and only two other spectra potentially
have host-galaxy contamination.  For SN~2001co, the continuum flux is
low enough that the choice of continuum does not result in any
difference in velocity shift.  For SN~2003dr, both the elliptical and
Sc templates result in a systematic blueshift of 80~\kms.

While it is possible that our choice of continuum is biasing the
measured velocity shifts for SNe~2003dg and 2003dr, we choose to use
the results from having a linear continuum.  This choice is 
conservative given our results (the other choices strengthen our
findings).  It also provides an easier path to reproducing the
methodology with future datasets.

Following the method of \citetalias{Perets10:05e} (and references
therein), we also attempted to subtract a ``photospheric'' SN
continuum from each spectrum assuming that the underlying continuum
is caused by photospheric SN emission.  We matched our spectra to
those of other SNe~Ib with similar spectral features away from the
nebular lines.  We then used these spectra as the underlying
continuum.  Doing so did not change the measured velocity in any
case.  For instance, performing the same continuum subtraction used by
\citetalias{Perets10:05e} for SN~2005E, the measured velocity offset
shifted by $<$1 pixel.

This is not surprising given the relative strength of the nebular
lines to the continuum.  In the case of SN~2005E, which has one of the
strongest continua, the continuum is only \about 15\% the flux of the
continuum-subtracted peak [\ion{Ca}{II}] flux.  Therefore, even 20\%
changes in the continuum level result in only a \about 3\% change in
the flux at any given wavelength.

Additionally, we examined the consequences of phase on the velocity
shift.  Unfortunately, only one Ca-rich SN has sufficient data to
examine any evolution.  SN~2010et has two nebular spectra of
reasonable quality that are separated by more than a week.  At epochs
of roughly 62 and 87~days after peak brightness, we measure a velocity
shift of $-330 \pm 60$ and $-430 \pm 60$~\kms, respectively.  These
measurements differ by only 1.1~$\sigma$, but may represent a slight
velocity gradient of $-4 \pm 3$~\kms\,day$^{-1}$.

SNe~Ia have been shown that their velocity shifts get redder with time
\citep[e.g.,][]{Silverman13}, eventually plateauing at a particular
velocity as the ejecta become optically thin.  While this is a
possibility for Ca-rich SNe, we note that SN~2012hn, which has the
largest velocity offset and is blueshifted also has one of the latest
spectra \citep[+150~days;][]{Valenti14}.

Currently, there are not enough data to determine the velocity
evolution of Ca-rich SNe with phase.  Nonetheless, we note that the
phase of a SN's spectrum and its galactic offset are uncorrelated
(correlation coefficient of $r = 0.006$), suggesting any potential
correlation between offset and velocity is not caused by a correlation
between phase and offset.

\subsection{Measured Velocity Shifts}

\begin{figure}
\begin{center}
\includegraphics[angle=90,width=3.12in]{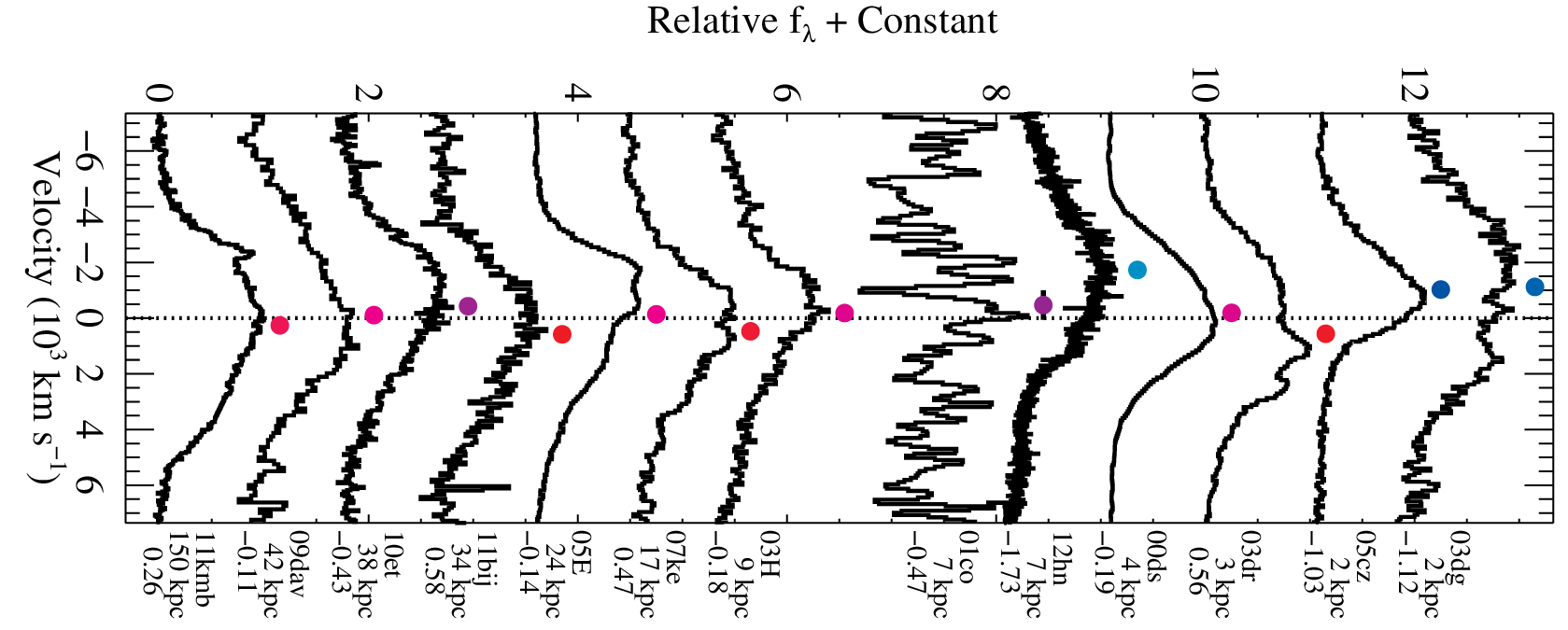}
\caption{Continuum-subtracted [\ion{Ca}{II}] $\lambda\lambda7291$, 7324
  profiles of the Ca-rich SN sample on a velocity scale relative to
  the host-galaxy frame.  The spectra are ordered by physical
  projected offset.  A circle indicates the emission-weighted velocity
  of each feature.  The uncertainty for each velocity measure is
  plotted, however, most uncertainties are smaller than the plotted
  symbols.  The offset and velocity (in $10^{3}$~km~s$^{-1}$) are
  listed next to each spectrum.}\label{f:ca2}
\end{center}
\end{figure}

In Figure~\ref{f:ca2}, we present the continuum-subtracted
[\ion{Ca}{II}] $\lambda\lambda7291$, 7324 profiles for the full Ca-rich
SN sample.  There is a large diversity to the line profiles with
roughly equal numbers having symmetric, skewed blue, and skewed red
profiles.

For each spectrum in Figure~\ref{f:ca2}, we mark the emission-weighted
velocity.  We present these data in Table~\ref{t:host}.

For four SNe (2005E, 2010et, PTF09dav, and PTF11bij),
\citetalias{Kasliwal12} measured [\ion{Ca}{II}] velocity shifts.
Their measurements deviate from ours by $-80$~\kms\ on average, with a
median difference of $-69$~\kms.  The largest difference is for
PTF09dav, where we measure a shift of $-110 \pm 60$~\kms, while
\citetalias{Kasliwal12} measure 250~\kms.  While
\citetalias{Kasliwal12} do not describe their method for measuring the
velocity shift or list any uncertainties, this represents at most a
5.9-$\sigma$ difference (using only the uncertainty associated with
our measurement).  All other differences are $<$200~\kms\ indicating a
general agreement in these measurements.

For 9 Ca-rich SNe, we could also measure an [\ion{O}{I}] velocity
shift, albeit with much larger uncertainties.  Performing a Bayesian
Monte-Carlo linear regression on the [\ion{Ca}{II}] and [\ion{O}{I}]
velocity shifts \citep{Kelly07}, we measure a slope and offset of $0.9
\pm 0.9$ and $0.2 \pm 0.5$, respectively.  Therefore, velocity
measurements from the [\ion{O}{I}] feature are consistent with those
of [\ion{Ca}{II}], providing additional confidence that the
[\ion{Ca}{II}] velocity shifts measure any bulk offset to the ejecta.

Five of 13 Ca-rich SNe have absolute velocity shifts of
$<$300~km~s$^{-1}$ relative to their host galaxy rest frame, which are
consistent with coming from galactic motion.  However, 8 SNe have
larger shifts, up to 1700~km~s$^{-1}$ relative to their host galaxy
rest frame.  Such large velocity shifts must originate from either an
asymmetric SN explosion or extreme line-of-sight motion for the
progenitor.


\section{Analysis}\label{s:anal}

\subsection{The Velocity-Offset Correlation}\label{ss:veloff}

\begin{figure*}
\begin{center}
\includegraphics[angle=0,width=3.2in]{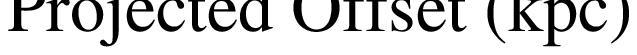}
\includegraphics[angle=0,width=3.2in]{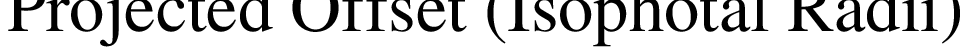}
\caption{Line-of-sight velocity shifts as a function of projected
  offset ({\it left}) and isophotal radial offset ({\it right}) for
  the Ca-rich sample.  Velocity uncertainties are typically smaller
  than the data points.  Only 4/13 SNe have a redshift, while only 1/6
  SNe within two isophotal radii (and 8~kpc) have a redshift and 3/7
  outside two isophotal radii (and 8~kpc) have
  redshifts.}\label{f:veloff}
\end{center}
\end{figure*}

For the Ca-rich SN sample, there is a large range of velocity shifts
(Section~\ref{s:shifts}).  However, this range changes with projected
offset.  SNe with small physical projected offsets have a much larger
range of velocity shifts than those with large projected offsets.  In
particular, within a projected offset of 8~kpc, 4/7 Ca-rich SNe have
absolute velocity shifts of $>$500~km~s$^{-1}$, while only 1/6 Ca-rich
SN beyond 8~kpc has such a large absolute velocity shift
(Figure~\ref{f:veloff}).

\begin{figure}
\begin{center}
\includegraphics[angle=0,width=3.2in]{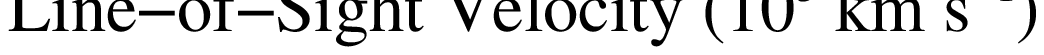}
\caption{Line-of-sight velocity shift CDFs for the Ca-rich SNe with a
  projected offset $<$8~kpc (black line) and $>$8 kpc (blue line).
  The SNe with small projected offsets have a wider distribution of
  line-of-sight velocity shifts ($p = 0.031$).}\label{f:velcdf}
\end{center}
\end{figure}

We also display these data as cumulative distribution functions (CDFs)
for Ca-rich SNe within and outside a projected offset of 8~kpc
(Figure~\ref{f:velcdf}) which also shows the different line-of-sight
velocity distributions for these subsamples.  To examine the
significance of this difference, we employ the Anderson-Darling
statistical test.  The Anderson-Darling test is similar to the
Komogorov-Smirnov test except it is more sensitive to differences in
the tails of distributions.  Such a test is particularly useful for
this analysis since we would naively expect any velocity shift
distribution to be centered at roughly zero velocity and differences
to be in the tails of the distributions.  If the distributions are
normal, the Anderson-Darling test reduces to the Komogorov-Smirnov
test.

The Anderson-Darling statistic for these two subsamples results in a
$p$-value of 0.031, indicating that the Ca-rich SNe with small and
large projected offsets are drawn from parent populations with
different velocity shifts.  This distinction is relatively insensitive
to the exact distance used to separate the subsamples.  Any chosen
separation ranging from 7.1 -- 16.7~kpc produces significantly
different subsamples.

We also note that while we may be systematically missing Ca-rich SNe
near the centers of their host galaxies with redshifted velocity
shifts (Section~\ref{ss:target}), the inclusion of these SNe, if they
have a similarly extreme velocity distribution as those with
blueshifted velocity shifts, would only make the tails of the
distributions more discrepant and difference between the subsamples
more significant.

This result is incredibly indicative.  First, if the velocity shifts
were primarily caused by the explosion or a binary orbital velocity,
there should be no correlation with projected offset.  Therefore, the
velocity shifts are almost certainly the result of extreme
line-of-sight motion for the progenitor.  To account for the SNe with
extreme velocity offsets, some progenitor systems must be given a
velocity kick of $\gtrsim$1500~\kms.

There is an additional clue which indicates that the velocity shifts
are not caused by the explosion.  Velocity shifts are seen in SNe~Ia
\citep[e.g.,][]{Maeda10:neb} and those shifts are thought to be a bulk
offset of the core of the ejecta relative to the outer layers.
However, SN~Ia explosions appear to be constrained to have relatively
small velocity shifts compared to the full-width at half-maximum
(FWHM) of their emission lines.  Using a sample of 22 SNe~Ia
\citep{Blondin12}, we find that the most extreme velocity shifts are
$<$20\% that of their FWHM.  For the Ca-rich SN sample, 3/13 have more
extreme velocity shifts (relative to FWHM) than any SN~Ia in the
\citet{Blondin12} sample.

\begin{figure}
\begin{center}
\includegraphics[angle=90,width=3.2in]{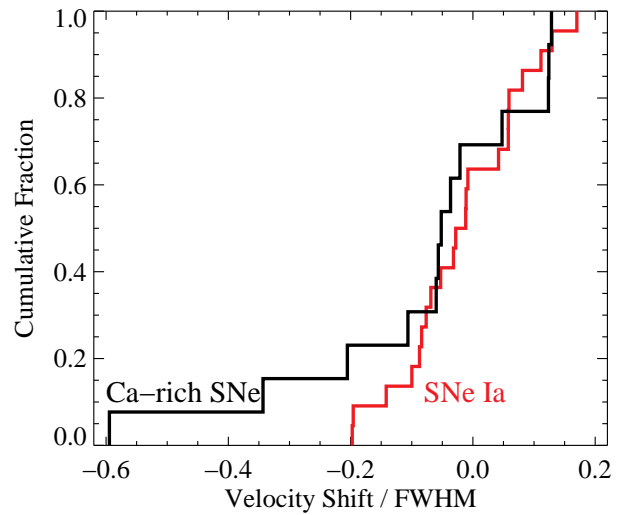}
\caption{CDFs of velocity shifts relative to nebular line FWHMs for
  the Ca-rich (black) and SN~Ia (red) samples.  Some Ca-rich SNe have
  much larger velocity shifts relative to the widths of their lines
  compared to the SN~Ia sample.  This indicates an additional velocity
  component.  The Anderson-Darling statistic for these two subsamples
  results in a $p$-value of 0.040.}\label{f:fwhm}
\end{center}
\end{figure}

Figure~\ref{f:fwhm} shows the CDFs of velocity shifts relative to
FWHMs for the Ca-rich and SN~Ia samples.  The Anderson-Darling
statistic for these two subsamples results in a $p$-value of 0.040,
indicating that these samples are drawn from different parent
populations.  This result implies that either Ca-rich SNe come from
significantly more asymmetric explosions than SNe~Ia or there is an
additional line-of-sight velocity component for some Ca-rich SNe.  The
latter is consistent with a correlation between velocity shift and
projected offset.

The velocity-offset correlation further indicates that the
line-of-sight motion for the progenitor must be correlated with the
projected offset.  The most obvious explanation is that most
progenitor systems originate near the centers of their host galaxies,
are given a large kick, and when they ultimately explode, have the
velocity kick imprinted in the kinematics of the ejecta.  For a
progenitor that travels along our line of sight, there will be no
projected offset for the SN, but its line-of-sight velocity offset
will be large.  Meanwhile, for a progenitor that travels perpendicular
to our line of sight, the SN will have a large projected offset but a
velocity offset close to zero (although the SN explosion or orbital
motion of the progenitor system, as well as measurement errors may
result in some velocity offset).

Therefore, Ca-rich SNe progenitor systems do {\it not} typically
originate from globular clusters, in-falling dwarf galaxies, the
intracluster medium, or other large-offset components of a galaxy.

From the velocity-offset data alone, it is possible that Ca-rich SN
progenitor systems originate in the disk or halo, offset from the
galaxy's nucleus, and (in some cases) are kicked to large offsets.

\subsection{An Excess of Blueshifts}\label{ss:blue}

The velocity offset distribution for the Ca-rich SN sample is skewed
to the blue.  Nine of the 13 SNe are blueshifted, while the remaining
4 are redshifted.  This is particularly striking when examining the 6
SNe within two isophotal radii of their host galaxies
(Figure~\ref{f:veloff}).  Only a single SN in this subsample is
redshifted.  However, 3/7 of the Ca-rich SNe with larger offsets are
redshifted.  The likelihood of having such skewed distributions are
only 17.5\% and 18.8\% for the full sample and small-offset subsample,
respectively.  This calculation accounts for the possibility that the
distribution could be skewed in the opposite direction as well (if
only considering blueshifts, the likelihood is half of that reported
above).

Although this is not a statistically significant result, it is
noteworthy because of a physically motivated reason for such a
difference.  If the progenitor systems were kicked from the centers of
their galaxies with velocities larger than the escape velocity of the
galaxy, then the velocity shift will be blueshifted for SNe on the
near side of the galaxy.  In such a scenario, redshifted SNe will be
on the far side of their host galaxies, and therefore will be more
likely to suffer from dust extinction.  Thus, Ca-rich SNe with
redshifted velocity shifts will be harder to detect, resulting in
under-representation in the full sample.

This is further support for most Ca-rich SN progenitors originating
from near the centers of their host galaxies.  If their progenitors
came from bound orbits (such as from a globular cluster population),
there should be an equal number of blueshifted and redshifted SNe on
the near side of the galaxy, and therefore no noticeable difference in
the observed population.  Furthermore, for progenitors in-falling
towards the host galaxy for the first time, we would expect more {\it
  redshifted} SNe, the opposite of what is observed.

\subsection{The Offset Angle Distribution}\label{ss:angle}

A final observational indication of the origin of Ca-rich SNe from the
SNe themselves is the angle between the SN and the disk of the galaxy,
which we call the offset angle.  A non-disk origin should have a flat
offset angle distribution (all angles are equally likely).  However, a
disk origin will tend to have more SNe with small angles relative to
the disk.  The exact offset angle distribution for the disk depends on
the average distance traveled by the progenitor system, $d$, relative
to the major axis of the disk, $r$.  For example, if $d/r = 0$,
corresponding to the SNe exploding in the disk, then all offset angles
would be zero.  A larger value of $d/r$ means that SNe will be further
above the disk (on average) and thus the offset angle distribution
will be flatter.

\begin{figure}
\begin{center}
\includegraphics[angle=90,width=3.2in]{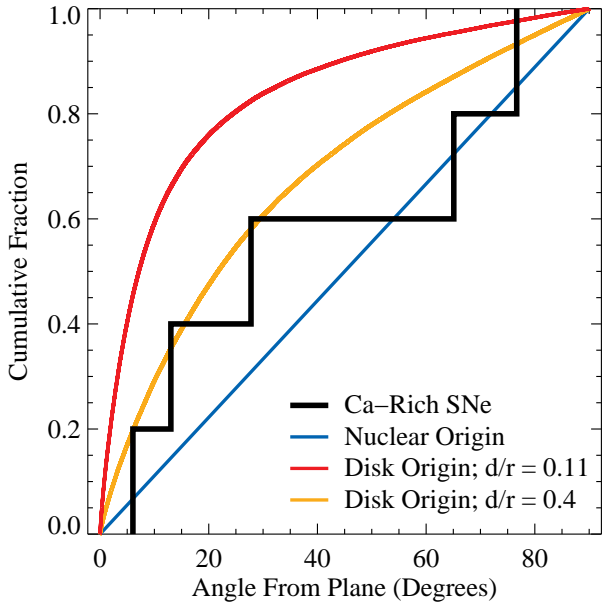}
\caption{Offset-angle CDF for Ca-rich SNe in edge-on galaxies (black).
  Also shown are theoretical distributions corresponding to a nuclear
  origin (blue line), a disk origin with the distance traveled by the
  progenitor system, $d$, relative to the major axis of the disk, $r$,
  of 0.11 or 0.4 ($d/r$; red and gold curves, respectively).  The disk
  origin with $d/r = 0.11$ is inconsistent with the observed
  distribution, while larger offsets relative to the height or the
  nuclear origin are consistent with the data.}\label{f:angle}
\end{center}
\end{figure}

In Figure~\ref{f:angle}, we plot the offset angle distribution for the
Ca-rich SNe with close to edge-on disks (axis ratios of $<$0.5).  This
corresponds to 5 SNe: SNe~2000ds, 2001co, 2003dg, 2003dr, and 2005E.
Of these 5, SNe~2000ds and 2003dr have offset angles $>$65$^{\circ}$.
The observed distribution is consistent with a flat distribution.

To generate the theoretical disk distributions, we assumed an
infinitely thin exponential disk perfectly edge on.  For this stellar
distribution, we chose random angles in three dimensions, and offset
an amount $d$ along those angles.  We then measured the offset angle
to the SN site.  As expected, a smaller value of $d/r$ results in a
distribution more peaked to zero degrees.

We use the Anderson-Darling statistic to determine what values of
$d/r$ are consistent with the data.  Ca-rich SNe are inconsistent with
coming from a disk population with ratios of $d/r \le 0.11$.

We cannot know the value of $d$ for our SNe, however, we can measure
the distance the SNe are above the disk, $z$, which must be a lower
limit on $d$.  For the edge-on subsample, there is a range of $0.04
\le z/r \le 0.56$.  Therefore, a value of $d/r \approx 0.4$ may be
appropriate, making a disk origin consistent with the observed angle
offset distribution.

More observations should further constrain any possible disk
contribution.


\section{A Basic Model}\label{s:model}

Previous studies have shown that Ca-rich SNe have He in their ejecta
(e.g., \citealt{Kawabata10}; \citetalias{Perets10:05e}) and come from
older stellar populations \citepalias[e.g.,][]{Perets10:05e}.  These
observations suggest an old progenitor system, and a massive-star
progenitor is unlikely.  The most obvious long-lived progenitor system
with a large abundance of helium is one that contains a He WD or a
hybrid He--C/O WD.  As these stars are not expected to explode in
isolation, a favorable progenitor system is a He--C/O WD binary
system.  Simulations suggest that a merger (or surface detonation
after mass transfer) for these systems can broadly reproduce the
Ca-rich SN properties \citep[e.g.,][]{Kromer10, Shen10, Waldman11,
  Sim12, Dessart15}.

Ca-rich SNe have an extended projected offset distribution
\citepalias[e.g.,][]{Kasliwal12}.  While some have attempted to
explain this as Ca-rich SNe having dwarf galaxy, globular cluster,
and/or intracluster origins, these scenarios are inconsistent with
other data.  For instance, if there is a dwarf galaxy origin, there
should not be such a strong correlation with nearby early-type
galaxies.

The strongest evidence against these previous scenarios is that there
is no indication of any stellar population at the position of any
Ca-rich SN to quite deep limits (\citetalias{Perets10:05e};
\citealt{Perets11}; \citetalias{Kasliwal12}; \citealt{Lyman13,
  Lyman14}).  This last observation requires that the progenitors of
Ca-rich SNe be born far from where they explode \citep{Lyman14}.

Our kinematic data indicates that the progenitors of Ca-rich SNe
originate near the centers of their galaxies.  There is a strong
correlation between the line-of-sight velocity shifts and projected
offsets for the Ca-rich SN sample (Section~\ref{ss:veloff}).  The
measured velocity shifts for some SNe is much larger than typical
stellar velocities and even larger than the escape velocity of a
galaxy (Section~\ref{s:shifts}).  In fact, the line-of-sight
velocities of several Ca-rich SNe is larger than the expected maximum
velocity from a disk origin \citep[\about 1000~km~s$^{-1}$;
e.g.,][]{Tauris15}, but is consistent with that of Galactic
hyper-velocity stars \citep[HVSs;][]{Brown14, Palladino14}.  From
kinematic data alone, it is most likely that Ca-rich progenitor
systems have been kicked after an interaction with a supermassive
black hole (SMBH) at the center of their host galaxies.

We have shown that almost all Ca-rich SN host galaxies are merging,
have recently merged, have characteristics indicative of a recent
merger, and/or are in dense environments where the probability of a
merger is high (Section~\ref{ss:merger}).  A binary SMBH can easily
fill its loss cone, and therefore the HVS ejection rate of such
systems are enhanced by a factor of \about $10^{4}$ \citep{Yu03} over
those of single SMBHs \citep[e.g.,][]{Hills88}.

The ejection rate of hyper-velocity binary stars (HVBSs) is expected
to be exceedingly low for single SMBHs, but perhaps half of all binary
stars interacting with a binary SMBH are ejected at hyper velocities
with the binary intact \citep{Lu07, Sesana09}.  This mechanism is also
expected to change the orbital properties of the binary.  In
particular, eccentricities of 0.9 -- 1 can be excited in these systems
\citep{Lu07}.  Presumably, such orbits will again circularize, but in
doing so, significantly reduce their separation.

If a close encounter with a SMBH is necessary to produce a Ca-rich SN
progenitor system or to significantly reduce its delay time, a recent
galaxy merger should also significantly enhance the Ca-rich SN rate.
In retrospect, one might have been able to predict that Ca-rich SNe
should be preferentially hosted by galaxies that have undergone recent
mergers.

Additionally, we have shown that a large fraction of Ca-rich SN host
galaxies also host an AGN (Section~\ref{ss:agn}).  After a merger, it
is expected that many SMBHs will have their gas supplies replenished,
and such galaxies will host an AGN.  However, such activity can be
relatively short-lived.  The typical duty cycle of an AGN is \about $4
\times 10^{7}$~years.  Because AGN activity is relatively short-lived,
any correlation between Ca-rich SNe and AGN activity suggests a
physical connection.  If one simply wants information to be
transferred between the nucleus and SN location, that information must
travel at a speed of $\gtrsim$750~\kms\ for a separation of 30~kpc
(similar to the separation between PTF11bij and the AGN in its host
galaxy, 34~kpc).  Notably, this velocity is larger than typical SN
kicks.  However, the expected hyper-velocity kicks from a SMBH
interaction is typically, \about 1500~\kms, larger than, but within a
factor of 2 of, the minimum speed for information travel in some
Ca-rich galactic systems.

With these strong pieces of evidence (helium in the system, an old
stellar population, a lack of a stellar population at the position of
the SNe, a correlation between line-of-sight velocity shifts and
projected offset, a significant number of host galaxies with a
high-likelihood of recent mergers, and a high AGN fraction for the
host galaxies), we suggest a possible progenitor system for Ca-rich
SNe: a He--C/O WD binary system that has recently been ejected from
its host galaxy after an encounter with a SMBH.  The ejection rate
will be significantly enhanced in galaxies with a binary SMBH, and
therefore also in galaxies which have recently undergone a merger.
Although it is possible for a Ca-rich SN to occur without an encounter
with a SMBH, this must not be the dominant progenitor path for Ca-rich
SNe.  If Ca-rich SNe commonly came from systems that did not interact
with a SMBH, we would not see the strong connections to mergers, AGN
activity, or even the large offsets.  Instead, the interaction with
the SMBH must significantly enhance the SN rate, likely by
significantly decreasing the delay time (perhaps from a typical delay
time of greater than a Hubble time to one similar to the AGN duty
cycle).

\subsection{Additional Model Constraints}

Such a progenitor model has other observational consequences
consistent with current data.  For instance, we would expect a flat
offset angle distribution.  Such a distribution is consistent with our
observations (Section~\ref{ss:angle}).  Additionally, as the SNe would
be on unbound orbits, the SNe projected on top of their galaxies would
have the highest velocity shifts.  However, if there is dust reddening
in the galaxy, we would expect redshifted SNe to be more difficult to
detect, resulting in an excess of SNe with blueshifted velocity
shifts.  Such an excess has also been observed
(Section~\ref{ss:blue}).

A final testable prediction of our model is that there should be an
anti-correlation between the magnitude of the velocity shift and the
size of the galaxy.  Larger galaxies with larger gravitational
potential wells should reduce the velocity shift more than smaller
galaxies.  Alternatively, if the SN progenitors originated in
satellite galaxies or globular clusters, the SN systemic velocity
relative to their host galaxy should increase with increasing
host-galaxy mass.

\begin{figure}
\begin{center}
\includegraphics[angle=0,width=3.2in]{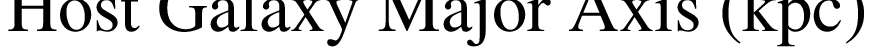}
\caption{Magnitude of velocity shifts as a function of host-galaxy
  size (as measured by the major axis) for the Ca-rich
  sample.}\label{f:size}
\end{center}
\end{figure}

Figure~\ref{f:size} shows the magnitude of the velocity shifts as a
function of host galaxy size for our sample.  There is an obvious
anti-correlation with the highest-velocity SNe coming from the most
compact galaxies.  While this is not a strong correlation ($r =
-0.29$), the size of a galaxy is not a perfect proxy for gravitational
potential, and cluster environments must further augment that
potential, this is further observational support for our model and
evidence against a dwarf galaxy or globular cluster origin.

\subsection{Model Implications and Predictions}

Our model requires that the inherent Ca-rich SN rate be extremely low
and an encounter with a SMBH must somehow substantially increase the
rate.  This is perhaps through changing the orbital dynamics of the
system or by triggering a physical change in a star in the system.
There is a clear example of the former possibility.  If the typical
progenitor system is a binary system with a separation large enough
such that a merger through gravitational radiation is longer than a
Hubble time, then the inherent Ca-rich SN rate will be necessarily
low.  However, an interaction with a SMBH (perhaps preceded with
other interactions in the nuclear star cluster) could then reduce the
merger time to significantly less than a Hubble time.

We can estimate the post-kick merger time for these systems.  If we
assume that a typical kick velocity is 1000~\kms, a star system would
have traveled for 30 or 150~Myr to travel 30 or 150~kpc, respectively.
While it is possible that some kicks result in bound orbits or that
some SNe with large projected offsets also have relatively large
line-of-sight offsets, these scenarios cannot be a significant portion
of the population.  Therefore, in our model, the typical post-kick
delay time is about 10--100~Myr.  This timescale is similar to that of
the AGN duty cycle.

However, if this delay time is equivalent to a binary merger time, it
must require a change in orbital separation by a factor of 10--100.
This requirement may be reduced some if the binary is initially
hardened by interactions in the nuclear star cluster.  If this is the
case, we would expect many more Ca-rich SNe in galactic nuclei.
Although there are known systematics for detecting nuclear SNe, the
lack of a substantial population of nuclear events in the current
sample is a hurdle for such an explanation.

Detailed calculations of the orbital parameters of potential
progenitor systems after ejection are required to measure a change in
final separation and merger time.

Another implication of the model is the relative rate of Ca-rich SNe
in galaxies with and without recent mergers.  Assuming that the
progenitor systems are formed before a merger and all galaxies have a
similar number of progenitor systems near their SMBHs, the relative
Ca-rich SN rate per unit mass should be equivalent to the relative
ejection rate, or \about $10^{4}$.  Since the ratio of recent mergers
to non-merging systems is roughly $10^{-2}$, the observed relative
rate should be \about $10^{2}$.  That is, for every Ca-rich SN in a
non-merging system, there should be 100 Ca-rich SNe in merging
systems.  The current Ca-rich sample appears to have \about 1/13
non-merging systems is consistent with this estimate.  However, this
estimate is only approximate and could be off by an order of
magnitude.  Similarly, the current sample may be highly biased or we
may have incorrectly estimated the number of merging systems.  This
question should be revisited with more detailed calculations and a
larger SN sample.

Nonetheless, the relative rate of Ca-rich SNe to SNe~Ia has been
measured to be \about 5\%.  If we assume that Ca-rich SNe occur
exclusively in merging systems, the Ca-rich SN rate must be \about 5
times that of the SN~Ia rate in these galaxies.  The SN~Ia rate in S0
galaxies is roughly 0.1 SNe~Ia per century \citep{Li11:rate3} or
$10^{-3}$~year$^{-1}$.  A $6.5 \times 10^{6} M_{\sun}$ SMBH with a
full loss cone has an ejection rate of \about 2 stars per year.
Therefore, Ca-rich SN progenitor systems must represent \about 0.25\%
of all ejected systems in these galaxies.  Considering the potentially
high survival rate of compact binary systems with binary SMBHs
\citep{Lu07}, this order-of-magnitude calculation appears to have the
Ca-rich SN rate compatible with the ejection rate.


\section{Discussion and Conclusions}\label{s:disc}

Ca-rich SNe are rare and peculiar low-luminosity Type I SNe with He in
their spectra near maximum brightness and relatively strong Ca
emission at late times.  While many of these SNe are within one
isophotal radius of their host galaxies, a significant percentage occur
far from their host galaxies --- some farther than any other detected
SN.

The Ca-rich SNe with small projected offsets tend to have much larger
line-of-sight velocity shifts as determine from the emission-weighted
velocity of forbidden lines.  The correlation between projected offset
and line-of-sight velocity shifts indicates that the progenitors of
Ca-rich SNe have been kicked from near the centers of their galaxies.

The highest velocity shifts are too large for a disk origin, but are
consistent with a kick from an interaction with a SMBH.  A nuclear
origin is further supported by a slight excess of blueshifted velocity
shifts with small offsets, an offset angle distribution consistent
with isotropic kicks from a galactic nucleus, and an anti-correlation
between the magnitude of the line-of-sight velocity shifts and galaxy
size.

The host galaxy population has a significant number of
merging/disturbed galaxies, S0 galaxies, and brightest cluster/group
galaxies --- all of which are more likely to have had recent mergers
than typical galaxies.  Such an observation must be linked to the
prevalence of Ca-rich SN progenitor systems.  A recent galaxy merger
enhancing the interaction rate between Ca-rich SN progenitor systems
and a SMBH is one possible reason for this observation.  Such a
scenario is expected to boost the SMBH ejection rate by a factor of
\about $10^{4}$, making observed and theoretical rates broadly
consistent.

The fraction of Ca-rich SN host galaxies with AGN is high.  The AGN
fraction is yet another indication of recent mergers.  The timescale
for AGN activity also places a soft upper limit on the delay time
since galaxy merger of \about $4 \times 10^{7}$~years and an
independent lower limit on the velocity of ejected systems.

The host-galaxy morphology distribution is indicative of an older
progenitor population \citepalias{Perets10:05e}.  The He in the
early-time spectra of Ca-rich SNe suggests a significant amount of He
is in the progenitor system.  As the progenitor system is likely old,
the source is likely a He WD.  

As a possible progenitor hypothesis, we propose the following: the
progenitor system for a Ca-rich SN is a binary system consisting of a
He WD with a C/O WD.  This system must have a typical merger time
longer than a Hubble time, which is most likely achieved with a
low-mass double WD system.  This system is kicked to high velocity
after interaction with a SMBH in the center of its host galaxy.
During this interaction, the binary is significantly hardened such
that the delay time is reduced to \about 50~Myr.  Such a possibility
seems plausible for binary systems ejected after interacting with a
binary SMBH, which is also expected to significantly reduce the merger
time \citep{Lu07}.  After traveling, on average, several to tens of
kpc, the system merges and results in a Ca-rich SN.

\citetalias{Perets10:05e} examined the possibility of the progenitor
of SN~2005E being ejected as a HVS, however they only examined this
for a potential high-mass ($M > 8 M_{\sun}$) progenitor.  They found
that such a progenitor system was unlikely, but did not address the
possibility of a low-mass binary progenitor ejected at hyper velocity.

If this scenario is correct, the progenitor systems of Ca-rich SNe
represent the first indication of extragalactic HVSs and the first
detection of HVBSs ejected from a galactic center.  These peculiar SNe
would provide the best way to observe HVSs beyond the local group.
Ca-rich SNe may be an excellent tracer of the SMBH population in
general and binary SMBHs in particular.  Ca-rich SN host galaxies may
provide an excellent input catalog for future binary SMBH studies.
Moreover, the Ca-rich SN rate may be an indirect tracer of the galaxy
merger rate.

We caution that the details of our toy model are currently poorly
constrained.  The current sample size is small and likely biased
against nuclear events (and perhaps against SNe with the largest
projected offsets).  An order of magnitude larger sample, which can be
achieved in the next few years with upcoming surveys combined with
dedicated spectroscopic resources, will provide necessary kinematic
constraints to infer the underlying properties of the Ca-rich SN
progenitor population.

\section*{Acknowledgments}

  {\it Facility:} Keck I(LRIS)

\bigskip

R.J.F.\ is incredibly thankful for extended discussions on this work
with P.\ Behroozi, and J.\ Guillochon, and E.\ Ramirez-Ruiz.  We also
thank L.\ Bildsten, R.\ Chornock, A.\ Filippenko, D.\ Kasen, K.\
Mandel, H.\ Perets, and K.\ Shen for discussing Ca-rich SNe and
providing insights on this work over the past several years.

We thank the participants of the ``Fast and Furious: Understanding
Exotic Astrophysical Transients'' workshop at the Aspen Center for
Physics, which is supported in part by the NSF under grant No.\
PHYS-1066293.  Multiple discussions at the workshop motivated portions
of this work.  R.J.F.\ also thanks the Aspen Center for Physics for
its hospitality during the ``Fast and Furious'' workshop in June 2014.

The Digitized Sky Surveys (DSS) were produced at the Space Telescope
Science Institute under U.S.\ Government grant NAG W--2166. The images
of these surveys are based on photographic data obtained using the
Oschin Schmidt Telescope on Palomar Mountain and the UK Schmidt
Telescope.  The plates were processed into the present compressed
digital form with the permission of these institutions.

This research has made extensive use of the NASA/IPAC Extragalactic
Database (NED) which is operated by the Jet Propulsion Laboratory,
California Institute of Technology, under contract with the National
Aeronautics and Space Administration.

This research has made use of the Keck Observatory Archive (KOA),
which is operated by the W. M. Keck Observatory and the NASA Exoplanet
Science Institute (NExScI), under contract with the National
Aeronautics and Space Administration.


\appendix
\section{Observations of PTF11\lowercase{kmb}}\label{a:11kmb}

PTF11kmb was discovered PTF on 24.24 August 2011 and was
spectroscopically classified as a SN~Ib from a spectrum obtained on 28
August 2011 \citep{Gal-Yam11}.

PTF11kmb was observed on 28.52 August 2011 (PI Filippenko; Program
U048LA) and 26.37 November 2011 (PI Kulkarni; Program C219LA) with the
Low Resolution Imaging Spectrometer \citep[LRIS;][]{Oke95} mounted on
the Keck~I telescope.  The August spectrum had a 450~s integration,
while the November spectrum had 1200~s blue-channel and $2 \times
570$~s red-channel integrations.  The August spectrum was originally
used to classify PTF11kmb as a SN~Ib.

We obtained these data through the Keck Observatory Archive.  Standard
CCD processing and spectrum extraction were accomplished with
IRAF\footnote{IRAF: the Image Reduction and Analysis Facility is
  distributed by the National Optical Astronomy Observatory, which is
  operated by the Association of Universities for Research in
  Astronomy, Inc.\ (AURA) under cooperative agreement with the
  National Science Foundation (NSF).}.  The data were extracted using
the optimal algorithm of \citet{Horne86}.  Low-order polynomial fits
to calibration-lamp spectra were used to establish the wavelength
scale, and small adjustments derived from night-sky lines in the
object frames were applied.  We employed our own IDL routines to flux
calibrate the data and remove telluric lines using the well-exposed
continua of spectrophotometric standards \citep{Wade88, Foley03,
  Silverman12:bsnip}.  The spectra are presented in
Figure~\ref{f:11kmb}.

\setcounter{figure}{11}
\begin{figure}
\begin{center}
\includegraphics[angle=90,width=3.2in]{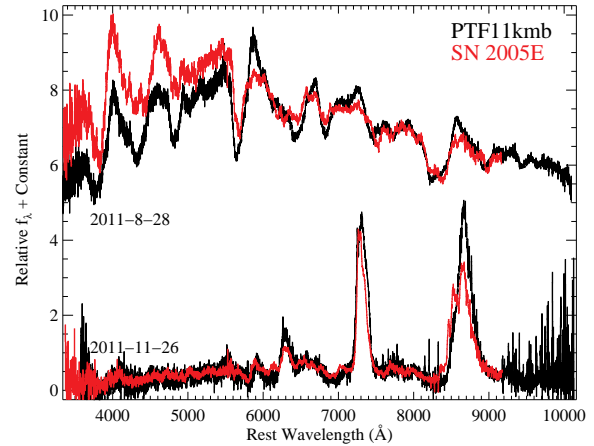}
\caption{Optical spectra of PTF11kmb (black curves) from 28 August
  2011 (top) and 26.37 November 2011 (bottom).  Comparison spectra of
  SN~2005E from 15.35 January 2005 (top) and 11.26 March 2005 (bottom;
  \citetalias{Perets10:05e}) are displayed in red.  All spectra have
  been corrected for Milky Way reddening.  The continuum is slightly
  different at early times, but this is likely the result of being at
  somewhat different phases or having different
  temperatures.}\label{f:11kmb}
\end{center}
\end{figure}

As the comparisons to SN~2005E in Figure~\ref{f:11kmb} show, PTF11kmb
is clearly a member of the Ca-rich class of SNe.

\bibliographystyle{mn2e}
\bibliography{../astro_refs}

\begin{thebibliography}{59}
\expandafter\ifx\csname natexlab\endcsname\relax\def\natexlab#1{#1}\fi

\bibitem[{{Baldwin} {et~al}\mbox{.}(1981){Baldwin}, {Phillips}, \&
  {Terlevich}}]{Baldwin81}
{Baldwin} J.~A., {Phillips} M.~M., {Terlevich} R., 1981, \pasp, 93, 5

\bibitem[{{Berger} {et~al}\mbox{.}(2013){Berger}, {Fong}, \&
  {Chornock}}]{Berger13:kilo}
{Berger} E., {Fong} W., {Chornock} R., 2013, \apjl, 774, L23

\bibitem[{{Blondin} {et~al}\mbox{.}(2012){Blondin}, {Matheson}, {Kirshner},
  {Mandel}, {Berlind}, {Calkins}, {Challis}, {Garnavich}, {Jha}, {Modjaz},
  {Riess}, \& {Schmidt}}]{Blondin12}
{Blondin} S. {et~al.}, 2012, \aj, 143, 126

\bibitem[{{Brown} {et~al}\mbox{.}(2014){Brown}, {Geller}, \&
  {Kenyon}}]{Brown14}
{Brown} W.~R., {Geller} M.~J., {Kenyon} S.~J., 2014, \apj, 787, 89

\bibitem[{{Dessart} \& {Hillier}(2015)}]{Dessart15}
{Dessart} L., {Hillier} D.~J., 2015, \mnras, 447, 1370

\bibitem[{{Drake} {et~al}\mbox{.}(2003){Drake}, {McGregor}, {Dopita}, \& {van
  Breugel}}]{Drake03}
{Drake} C.~L., {McGregor} P.~J., {Dopita} M.~A., {van Breugel} W.~J.~M., 2003,
  \aj, 126, 2237

\bibitem[{{Filippenko} {et~al}\mbox{.}(2003){Filippenko}, {Chornock}, {Swift},
  {Modjaz}, {Simcoe}, \& {Rauch}}]{Filippenko03:carich}
{Filippenko} A.~V., {Chornock} R., {Swift} B., {Modjaz} M., {Simcoe} R.,
  {Rauch} M., 2003, \iaucirc, 8159, 2

\bibitem[{{Filippenko} {et~al}\mbox{.}(2001){Filippenko}, {Li}, {Treffers}, \&
  {Modjaz}}]{Filippenko01}
{Filippenko} A.~V., {Li} W.~D., {Treffers} R.~R., {Modjaz} M., 2001, in ASP
  Conf. Ser. 246: IAU Colloq. 183: Small Telescope Astronomy on Global Scales,
  {Paczynski} B., {Chen} W.-P., {Lemme} C., eds., pp. 121--+

\bibitem[{{Foley} {et~al}\mbox{.}(2013){Foley}, {Challis}, {Chornock},
  {Ganeshalingam}, {Li}, {Marion}, {Morrell}, {Pignata}, {Stritzinger},
  {Silverman}, {Wang}, {Anderson}, {Filippenko}, {Freedman}, {Hamuy}, {Jha},
  {Kirshner}, {McCully}, {Persson}, {Phillips}, {Reichart}, \&
  {Soderberg}}]{Foley13:iax}
{Foley} R.~J. {et~al.}, 2013, \apj, 767, 57

\bibitem[{{Foley} {et~al}\mbox{.}(2010){Foley}, {Narayan}, {Challis},
  {Filippenko}, {Kirshner}, {Silverman}, \& {Steele}}]{Foley10:06bt}
{Foley} R.~J., {Narayan} G., {Challis} P.~J., {Filippenko} A.~V., {Kirshner}
  R.~P., {Silverman} J.~M., {Steele} T.~N., 2010, \apj, 708, 1748

\bibitem[{{Foley} {et~al}\mbox{.}(2003){Foley}, {Papenkova}, {Swift},
  {Filippenko}, {Li}, {Mazzali}, {Chornock}, {Leonard}, \& {Van Dyk}}]{Foley03}
{Foley} R.~J. {et~al.}, 2003, \pasp, 115, 1220

\bibitem[{{Gal-Yam} {et~al}\mbox{.}(2011){Gal-Yam}, {Xu}, {Ben-Ami}, {Arcavi},
  {Sternberg}, {Nugent}, {Cao}, {Konidaris}, {Levitan}, {Maguire}, {Pan},
  {Cenko}, {Silverman}, {Kandrashoff}, {Bloom}, {Walker}, \&
  {Groot}}]{Gal-Yam11}
{Gal-Yam} A. {et~al.}, 2011, The Astronomer's Telegram, 3631, 1

\bibitem[{{Graham} {et~al}\mbox{.}(2005){Graham}, {Li}, {Schwartz}, \&
  {Trondal}}]{Graham05}
{Graham} J., {Li} W., {Schwartz} M., {Trondal} O., 2005, \iaucirc, 8465, 1

\bibitem[{{Hills}(1988)}]{Hills88}
{Hills} J.~G., 1988, \nat, 331, 687

\bibitem[{{Ho}(2008)}]{Ho08}
{Ho} L.~C., 2008, \araa, 46, 475

\bibitem[{{Ho} {et~al}\mbox{.}(1997){Ho}, {Filippenko}, \&
  {Sargent}}]{Ho97:params}
{Ho} L.~C., {Filippenko} A.~V., {Sargent} W.~L.~W., 1997, \apjs, 112, 315

\bibitem[{{Horne}(1986)}]{Horne86}
{Horne} K., 1986, \pasp, 98, 609

\bibitem[{{Kasliwal} {et~al}\mbox{.}(2012){Kasliwal}, {Kulkarni}, {Gal-Yam},
  {Nugent}, {Sullivan}, {Bildsten}, {Yaron}, {Perets}, {Arcavi}, {Ben-Ami},
  {Bhalerao}, {Bloom}, {Cenko}, {Filippenko}, {Frail}, {Ganeshalingam},
  {Horesh}, {Howell}, {Law}, {Leonard}, {Li}, {Ofek}, {Polishook}, {Poznanski},
  {Quimby}, {Silverman}, {Sternberg}, \& {Xu}}]{Kasliwal12}
{Kasliwal} M.~M. {et~al.}, 2012, \apj, 755, 161

\bibitem[{{Kauffmann} {et~al}\mbox{.}(2003){Kauffmann}, {Heckman}, {Tremonti},
  {Brinchmann}, {Charlot}, {White}, {Ridgway}, {Brinkmann}, {Fukugita}, {Hall},
  {Ivezi{\'c}}, {Richards}, \& {Schneider}}]{Kauffmann03}
{Kauffmann} G. {et~al.}, 2003, \mnras, 346, 1055

\bibitem[{{Kaufman} {et~al}\mbox{.}(2012){Kaufman}, {Grupe}, {Elmegreen},
  {Elmegreen}, {Struck}, \& {Brinks}}]{Kaufman12}
{Kaufman} M., {Grupe} D., {Elmegreen} B.~G., {Elmegreen} D.~M., {Struck} C.,
  {Brinks} E., 2012, \aj, 144, 156

\bibitem[{{Kawabata} {et~al}\mbox{.}(2010){Kawabata}, {Maeda}, {Nomoto},
  {Taubenberger}, {Tanaka}, {Deng}, {Pian}, {Hattori}, \&
  {Itagaki}}]{Kawabata10}
{Kawabata} K.~S. {et~al.}, 2010, \nat, 465, 326

\bibitem[{{Kelly}(2007)}]{Kelly07}
{Kelly} B.~C., 2007, \apj, 665, 1489

\bibitem[{{Kewley} {et~al}\mbox{.}(2001){Kewley}, {Dopita}, {Sutherland},
  {Heisler}, \& {Trevena}}]{Kewley01}
{Kewley} L.~J., {Dopita} M.~A., {Sutherland} R.~S., {Heisler} C.~A., {Trevena}
  J., 2001, \apj, 556, 121

\bibitem[{{Kewley} {et~al}\mbox{.}(2006){Kewley}, {Groves}, {Kauffmann}, \&
  {Heckman}}]{Kewley06}
{Kewley} L.~J., {Groves} B., {Kauffmann} G., {Heckman} T., 2006, \mnras, 372,
  961

\bibitem[{{Kromer} {et~al}\mbox{.}(2010){Kromer}, {Sim}, {Fink}, {R{\"o}pke},
  {Seitenzahl}, \& {Hillebrandt}}]{Kromer10}
{Kromer} M., {Sim} S.~A., {Fink} M., {R{\"o}pke} F.~K., {Seitenzahl} I.~R.,
  {Hillebrandt} W., 2010, \apj, 719, 1067

\bibitem[{{Leaman} {et~al}\mbox{.}(2011){Leaman}, {Li}, {Chornock}, \&
  {Filippenko}}]{Leaman11}
{Leaman} J., {Li} W., {Chornock} R., {Filippenko} A.~V., 2011, \mnras, 412,
  1419

\bibitem[{{Li} {et~al}\mbox{.}(2011){Li}, {Chornock}, {Leaman}, {Filippenko},
  {Poznanski}, {Wang}, {Ganeshalingam}, \& {Mannucci}}]{Li11:rate3}
{Li} W., {Chornock} R., {Leaman} J., {Filippenko} A.~V., {Poznanski} D., {Wang}
  X., {Ganeshalingam} M., {Mannucci} F., 2011, \mnras, 412, 1473

\bibitem[{{Liu} {et~al}\mbox{.}(2011){Liu}, {Shen}, {Strauss}, \&
  {Hao}}]{Liu11}
{Liu} X., {Shen} Y., {Strauss} M.~A., {Hao} L., 2011, \apj, 737, 101

\bibitem[{{Lu} {et~al}\mbox{.}(2007){Lu}, {Yu}, \& {Lin}}]{Lu07}
{Lu} Y., {Yu} Q., {Lin} D.~N.~C., 2007, \apjl, 666, L89

\bibitem[{{Lyman} {et~al}\mbox{.}(2013){Lyman}, {James}, {Perets}, {Anderson},
  {Gal-Yam}, {Mazzali}, \& {Percival}}]{Lyman13}
{Lyman} J.~D., {James} P.~A., {Perets} H.~B., {Anderson} J.~P., {Gal-Yam} A.,
  {Mazzali} P., {Percival} S.~M., 2013, \mnras, 434, 527

\bibitem[{{Lyman} {et~al}\mbox{.}(2014){Lyman}, {Levan}, {Church}, {Davies}, \&
  {Tanvir}}]{Lyman14}
{Lyman} J.~D., {Levan} A.~J., {Church} R.~P., {Davies} M.~B., {Tanvir} N.~R.,
  2014, \mnras, 444, 2157

\bibitem[{{Maeda} {et~al}\mbox{.}(2010){Maeda}, {Taubenberger}, {Sollerman},
  {Mazzali}, {Leloudas}, {Nomoto}, \& {Motohara}}]{Maeda10:neb}
{Maeda} K., {Taubenberger} S., {Sollerman} J., {Mazzali} P.~A., {Leloudas} G.,
  {Nomoto} K., {Motohara} K., 2010, \apj, 708, 1703

\bibitem[{{Moore} {et~al}\mbox{.}(1998){Moore}, {Lake}, \& {Katz}}]{Moore98}
{Moore} B., {Lake} G., {Katz} N., 1998, \apj, 495, 139

\bibitem[{{Moore} {et~al}\mbox{.}(1999){Moore}, {Lake}, {Quinn}, \&
  {Stadel}}]{Moore99}
{Moore} B., {Lake} G., {Quinn} T., {Stadel} J., 1999, \mnras, 304, 465

\bibitem[{{Nagar} {et~al}\mbox{.}(2005){Nagar}, {Falcke}, \&
  {Wilson}}]{Nagar05}
{Nagar} N.~M., {Falcke} H., {Wilson} A.~S., 2005, \aap, 435, 521

\bibitem[{{Oke} {et~al}\mbox{.}(1995){Oke}, {Cohen}, {Carr}, {Cromer},
  {Dingizian}, {Harris}, {Labrecque}, {Lucinio}, {Schaal}, {Epps}, \&
  {Miller}}]{Oke95}
{Oke} J.~B. {et~al.}, 1995, \pasp, 107, 375

\bibitem[{{Palladino} {et~al}\mbox{.}(2014){Palladino}, {Schlesinger},
  {Holley-Bockelmann}, {Allende Prieto}, {Beers}, {Lee}, \&
  {Schneider}}]{Palladino14}
{Palladino} L.~E., {Schlesinger} K.~J., {Holley-Bockelmann} K., {Allende
  Prieto} C., {Beers} T.~C., {Lee} Y.~S., {Schneider} D.~P., 2014, \apj, 780, 7

\bibitem[{{Peletier} {et~al}\mbox{.}(1990){Peletier}, {Davies}, {Illingworth},
  {Davis}, \& {Cawson}}]{Peletier90}
{Peletier} R.~F., {Davies} R.~L., {Illingworth} G.~D., {Davis} L.~E., {Cawson}
  M., 1990, \aj, 100, 1091

\bibitem[{{Perets} {et~al}\mbox{.}(2011){Perets}, {Gal-yam}, {Crockett},
  {Anderson}, {James}, {Sullivan}, {Neill}, \& {Leonard}}]{Perets11}
{Perets} H.~B., {Gal-yam} A., {Crockett} R.~M., {Anderson} J.~P., {James}
  P.~A., {Sullivan} M., {Neill} J.~D., {Leonard} D.~C., 2011, \apjl, 728, L36

\bibitem[{{Perets} {et~al}\mbox{.}(2010){Perets}, {Gal-Yam}, {Mazzali},
  {Arnett}, {Kagan}, {Filippenko}, {Li}, {Arcavi}, {Cenko}, {Fox}, {Leonard},
  {Moon}, {Sand}, {Soderberg}, {Anderson}, {James}, {Foley}, {Ganeshalingam},
  {Ofek}, {Bildsten}, {Nelemans}, {Shen}, {Weinberg}, {Metzger}, {Piro},
  {Quataert}, {Kiewe}, \& {Poznanski}}]{Perets10:05e}
{Perets} H.~B. {et~al.}, 2010, \nat, 465, 322

\bibitem[{{Quimby} {et~al}\mbox{.}(2011){Quimby}, {Kulkarni}, {Kasliwal},
  {Gal-Yam}, {Arcavi}, {Sullivan}, {Nugent}, {Thomas}, {Howell}, {Nakar},
  {Bildsten}, {Theissen}, {Law}, {Dekany}, {Rahmer}, {Hale}, {Smith}, {Ofek},
  {Zolkower}, {Velur}, {Walters}, {Henning}, {Bui}, {McKenna}, {Poznanski},
  {Cenko}, \& {Levitan}}]{Quimby11}
{Quimby} R.~M. {et~al.}, 2011, \nat, 474, 487

\bibitem[{{Sesana} {et~al}\mbox{.}(2009){Sesana}, {Madau}, \&
  {Haardt}}]{Sesana09}
{Sesana} A., {Madau} P., {Haardt} F., 2009, \mnras, 392, L31

\bibitem[{{Shen} {et~al}\mbox{.}(2010){Shen}, {Kasen}, {Weinberg}, {Bildsten},
  \& {Scannapieco}}]{Shen10}
{Shen} K.~J., {Kasen} D., {Weinberg} N.~N., {Bildsten} L., {Scannapieco} E.,
  2010, \apj, 715, 767

\bibitem[{{Silverman} {et~al}\mbox{.}(2012){Silverman}, {Foley}, {Filippenko},
  {Ganeshalingam}, {Barth}, {Chornock}, {Griffith}, {Kong}, {Lee}, {Leonard},
  {Matheson}, {Miller}, {Steele}, {Barris}, {Bloom}, {Cobb}, {Coil},
  {Desroches}, {Gates}, {Ho}, {Jha}, {Kandrashoff}, {Li}, {Mandel}, {Modjaz},
  {Moore}, {Mostardi}, {Papenkova}, {Park}, {Perley}, {Poznanski}, {Reuter},
  {Scala}, {Serduke}, {Shields}, {Swift}, {Tonry}, {Van Dyk}, {Wang}, \&
  {Wong}}]{Silverman12:bsnip}
{Silverman} J.~M. {et~al.}, 2012, \mnras, 425, 1789

\bibitem[{{Silverman} {et~al}\mbox{.}(2013){Silverman}, {Ganeshalingam}, \&
  {Filippenko}}]{Silverman13}
{Silverman} J.~M., {Ganeshalingam} M., {Filippenko} A.~V., 2013, \mnras, 529

\bibitem[{{Sim} {et~al}\mbox{.}(2012){Sim}, {Fink}, {Kromer}, {R{\"o}pke},
  {Ruiter}, \& {Hillebrandt}}]{Sim12}
{Sim} S.~A., {Fink} M., {Kromer} M., {R{\"o}pke} F.~K., {Ruiter} A.~J.,
  {Hillebrandt} W., 2012, \mnras, 420, 3003

\bibitem[{{Smith} {et~al}\mbox{.}(2008){Smith}, {Foley}, \&
  {Filippenko}}]{Smith08:06jc}
{Smith} N., {Foley} R.~J., {Filippenko} A.~V., 2008, \apj, 680, 568

\bibitem[{{Smith} {et~al}\mbox{.}(2007){Smith}, {Li}, {Foley}, {Wheeler},
  {Pooley}, {Chornock}, {Filippenko}, {Silverman}, {Quimby}, {Bloom}, \&
  {Hansen}}]{Smith07:06gy}
{Smith} N. {et~al.}, 2007, \apj, 666, 1116

\bibitem[{{Sullivan} {et~al}\mbox{.}(2011){Sullivan}, {Kasliwal}, {Nugent},
  {Howell}, {Thomas}, {Ofek}, {Arcavi}, {Blake}, {Cooke}, {Gal-Yam}, {Hook},
  {Mazzali}, {Podsiadlowski}, {Quimby}, {Bildsten}, {Bloom}, {Cenko},
  {Kulkarni}, {Law}, \& {Poznanski}}]{Sullivan11:09dav}
{Sullivan} M. {et~al.}, 2011, \apj, 732, 118

\bibitem[{{Tanvir} {et~al}\mbox{.}(2013){Tanvir}, {Levan}, {Fruchter},
  {Hjorth}, {Hounsell}, {Wiersema}, \& {Tunnicliffe}}]{Tanvir13}
{Tanvir} N.~R., {Levan} A.~J., {Fruchter} A.~S., {Hjorth} J., {Hounsell} R.~A.,
  {Wiersema} K., {Tunnicliffe} R.~L., 2013, \nat, 500, 547

\bibitem[{{Tauris}(2015)}]{Tauris15}
{Tauris} T.~M., 2015, \mnras, 448, L6

\bibitem[{{Tremonti} {et~al}\mbox{.}(2004){Tremonti}, {Heckman}, {Kauffmann},
  {Brinchmann}, {Charlot}, {White}, {Seibert}, {Peng}, {Schlegel}, {Uomoto},
  {Fukugita}, \& {Brinkmann}}]{Tremonti04}
{Tremonti} C.~A. {et~al.}, 2004, \apj, 613, 898

\bibitem[{{Valenti} {et~al}\mbox{.}(2014){Valenti}, {Yuan}, {Taubenberger},
  {Maguire}, {Pastorello}, {Benetti}, {Smartt}, {Cappellaro}, {Howell},
  {Bildsten}, {Moore}, {Stritzinger}, {Anderson}, {Benitez-Herrera}, {Bufano},
  {Gonzalez-Gaitan}, {McCrum}, {Pignata}, {Fraser}, {Gal-Yam}, {Le Guillou},
  {Inserra}, {Reichart}, {Scalzo}, {Sullivan}, {Yaron}, \& {Young}}]{Valenti14}
{Valenti} S. {et~al.}, 2014, \mnras, 437, 1519

\bibitem[{{V{\'e}ron-Cetty} \& {V{\'e}ron}(2006)}]{Veron-Cetty06}
{V{\'e}ron-Cetty} M.-P., {V{\'e}ron} P., 2006, \aap, 455, 773

\bibitem[{{Wade} \& {Horne}(1988)}]{Wade88}
{Wade} R.~A., {Horne} K., 1988, \apj, 324, 411

\bibitem[{{Waldman} {et~al}\mbox{.}(2011){Waldman}, {Sauer}, {Livne}, {Perets},
  {Glasner}, {Mazzali}, {Truran}, \& {Gal-Yam}}]{Waldman11}
{Waldman} R., {Sauer} D., {Livne} E., {Perets} H., {Glasner} A., {Mazzali} P.,
  {Truran} J.~W., {Gal-Yam} A., 2011, \apj, 738, 21

\bibitem[{{Yaron} \& {Gal-Yam}(2012)}]{Yaron12}
{Yaron} O., {Gal-Yam} A., 2012, \pasp, 124, 668

\bibitem[{{Yu} \& {Tremaine}(2003)}]{Yu03}
{Yu} Q., {Tremaine} S., 2003, \apj, 599, 1129

\bibitem[{{Yuan} {et~al}\mbox{.}(2013){Yuan}, {Kobayashi}, {Schmidt},
  {Podsiadlowski}, {Sim}, \& {Scalzo}}]{Yuan13}
{Yuan} F., {Kobayashi} C., {Schmidt} B.~P., {Podsiadlowski} P., {Sim} S.~A.,
  {Scalzo} R.~A., 2013, \mnras, 432, 1680

\end{thebibliography}

\onecolumn
\begin{deluxetable}{llccccrcl}
\tabletypesize{\footnotesize}
\tablewidth{0pt}
\tablecaption{Ca-rich SN Host-galaxy Demographics\label{t:host}}
\tablehead{
\colhead{SN} &
\colhead{Host} &
\colhead{Morphology} &
\colhead{AGN} &
\colhead{$z$} &
\colhead{Scale} &
\colhead{Major} &
\colhead{Axis Ratio} &
\colhead{Galactic} \\
\colhead{ } &
\colhead{ } &
\colhead{ } &
\colhead{ } &
\colhead{ } &
\colhead{(kpc/$^{\prime\prime}$)} &
\colhead{Axis ($^{\prime\prime}$)} &
\colhead{($b/a$)} &
\colhead{P.A.\ ($^{\circ}$)}}

\startdata

2000ds   & NGC~2768                & E                 & LINER        & 0.0045 & 0.113 & 65.70 & 0.460 & $-87.5$            \\
2001co   & NGC~5559                & Sb (BGG)          & BPT Comp     & 0.0172 & 0.383 & 18.55 & 0.300 & \hphantom{$-$}60   \\
2003H    & NGC~2207                & Sbc (interacting) & AGN          & 0.0091 & 0.171 & 36.25 & 0.680 & \hphantom{$-$}70   \\
2003dg   & UGC~6934                & Scd               & \nodata      & 0.0183 & 0.405 & 11.15 & 0.320 & $-35$              \\
2003dr   & NGC~5714                & Scd (BGG)         & \nodata      & 0.0075 & 0.184 & 28.30 & 0.160 & \hphantom{$-$}80   \\
2005E    & NGC~1032                & S0/a              & Radio Excess & 0.0090 & 0.173 & 33.80 & 0.400 & \hphantom{$-$}70   \\
2005cz   & NGC~4589                & E (BGG)           & LINER        & 0.0066 & 0.158 & 34.75 & 0.750 & \hphantom{$-$}92.5 \\
2007ke   & NGC~1129                & E (BCG; merger)   & \nodata      & 0.0173 & 0.340 & 38.15 & 0.880 & \hphantom{$-$}70   \\
2010et   & CGCG~170-011            & E (group)         & BPT AGN      & 0.0233 & 0.495 &  7.00 & 0.800 & $-30$              \\
2012hn   & NGC~2272                & S0 (BGG)          & \nodata      & 0.0071 & 0.142 & 25.50 & 0.740 & $-55$              \\
PTF09dav & 2MASS~J22465295+2138221 & Disturbed Spiral  & \nodata      & 0.0371 & 0.735 &  4.35 & 0.740 & $-85$              \\
PTF11bij & IC~3956                 & S0? (BGG)         & AGN          & 0.0347 & 0.720 &  5.45 & 0.900 & $-20$              \\
PTF11kmb & NGC~7265                & S0 (BGG)          & \nodata      & 0.0170 & 0.343 & 26.25 & 0.760 & $-10$

\enddata
\end{deluxetable}

\begin{deluxetable}{lrrrrr}
\tabletypesize{\footnotesize}
\tablewidth{0pt}
\tablecaption{Ca-rich SN Sample Demographics\label{t:sn}}
\tablehead{
\colhead{SN} &
\colhead{SN Offset} &
\colhead{SN Offset} &
\colhead{SN Offset} &
\colhead{Angle} &
\colhead{Velocity Shift} \\
\colhead{ } &
\colhead{($^{\prime\prime}$)} &
\colhead{(kpc)} &
\colhead{(Isophotal Radii)} &
\colhead{Offset ($^{\circ}$)} &
\colhead{(km~s$^{-1}$)}}

\startdata

2000ds   &  33.4 &   3.77 &  1.08 & 76.67 & \hphantom{1}$-190$ \hphantom{1}(30) \\
2001co   &  18.5 &   7.07 &  1.22 & 13.03 & \hphantom{1}$-470$ (530) \\
2003H    &  51.0 &   8.73 &  1.52 & 22.47 & \hphantom{1}$-180$ (130) \\
2003dg   &   4.1 &   1.66 &  0.39 &  6.05 & $-1120$            \hphantom{1}(40) \\
2003dr   &  14.4 &   2.65 &  2.89 & 65.09 & \hphantom{$-1$}560 \hphantom{1}(40) \\
2005E    & 140.3 &  24.27 &  6.07 & 27.78 & \hphantom{1}$-140$ \hphantom{1}(30) \\
2005cz   &  13.4 &   2.12 &  0.41 & 24.07 & $-1030$            (140) \\
2007ke   &  49.2 &  16.71 &  1.31 & 17.31 & \hphantom{$-1$}470 (120) \\
2010et   &  76.0 &  37.64 & 10.95 &  9.44 & \hphantom{1}$-430$ \hphantom{1}(70) \\
2012hn   &  47.4 &   6.73 &  2.48 & 75.50 & $-1730$            \hphantom{1}(70) \\
PTF09dav &  56.8 &  41.75 & 16.09 & 52.29 & \hphantom{1}$-110$ \hphantom{1}(60) \\
PTF11bij &  47.8 &  34.42 &  8.80 & 10.00 & \hphantom{$-1$}580 (100) \\
PTF11kmb & 437.5 & 150.05 & 20.46 & 56.36 & \hphantom{$-1$}260 \hphantom{1}(20)

\enddata
\end{deluxetable}


\end{document}